\documentclass[11pt,a4paper]{article}

\usepackage{epsfig}
\usepackage{a4wide}
\usepackage{latexsym}
\usepackage{graphicx}
\usepackage{amsmath}
\usepackage{slashed}
\usepackage{amsfonts}
\usepackage{amssymb}
\usepackage{mathrsfs}
\usepackage{appendix}
\usepackage[usenames,dvips]{color}
\usepackage[colorlinks=true,urlcolor=PineGreen,anchorcolor=PineGreen,citecolor=PineGreen,filecolor=PineGreen,linkcolor=PineGreen,menucolor=PineGreen,pagecolor=PineGreen,linktocpage=true,pdfproducer=medialab]{hyperref}
\usepackage{layout}

\allowdisplaybreaks[4]
\numberwithin{equation}{section}

\begin{document}

\begin{titlepage}

\begin{center}
{\LARGE\textbf{Self-duality, helicity conservation and normal ordering in nonlinear QED}}\\[1.5 cm]
\bigskip

\textbf{ Ji\v{r}\'{\i} Novotn\'y}\footnote{
for email use: novotny at ipnp.troja.mff.cuni.cz},  \\[1 cm]
\textit{Institute of Particle and Nuclear Physics, Faculty of
Mathematics
and Physics,}\\[0pt]
\textit{Charles University, V Hole\v{s}ovi\v{c}k\'ach 2, CZ-180 00
Prague 8,
Czech Republic} \\[0.5 cm]
\end{center}

\begin{abstract}
We give a  proof of the equivalence of the electric-magnetic duality on  one side and helicity conservation of the tree level amplitudes  on the other  side within general models of nonlinear electrodynamics. Using modified Feynman rules derived from generalized normal ordered Lagrangian we discuss the interrelation of the above two properties of the theory  also at higher loops. As an illustration we present two explicit examples, namely we find  the generalized normal ordered Lagrangian for the Born-Infeld theory and derive a semi-closed expression for the Lagrangian of the  Bossard-Nicolai model (in terms of the weak field expansion with explicitly known coefficients) from its normal ordered form.
\end{abstract}
\end{titlepage}

\section{Introduction}

The electric-magnetic duality is a remarkable on shell symmetry of the
equation of motion of the Maxwell theory. It also holds for some of its
nonlinear generalization the most famous of which is the one constructed by
Born and Infeld in the thirties \cite{Born:1934gh}. The general aspects of
this type of duality and of its extensions were studied in detail by
Gaillard and Zumino in a seminal paper \cite{Gaillard:1981rj}, where also
the famous Noether-Gaillard-Zumino (NGZ) identity expressing the necessary
and sufficient condition for the Lagrangian of the duality invariant theory
was obtained. An iterative solution of the NGZ condition in terms of one
arbitrary function was found in \cite{Gibbons:1995cv} prooving at the same
time that Maxwell and BI theories are not the only self-dual cases and that
there is an infinite class of such theories. The general solution of NGZ
condition was found by by Gaillard and Zumino in \cite{Gaillard:1997rt}.
Their implicit construction of the solution is again parameterized by means
of one arbitrary function of one external variable. This variable is
implicitly determined as a solution of certain (in general case
transcendental) equation. Other alternative solution of the NGZ condition of
this type was constructed by Hatsuda, Kamimura and Sekiya in \cite%
{Hatsuda:1999ys}, where also non-trivial explicit examples of the self-dual
Lagrangians were given in a closed form. Different approach to construction
of the Lagrangian of self-dual theories, which generalized the
Bossard-Nicolai procedure \cite{Bossard:2011ij}, was elaborated by Carrasco,
Kallosh and Roiban in \cite{Carrasco:2011jv} and solutions of the
corresponding nonlinear twisted self-duality constraints for the Mawell
case, Born-Infeld (BI) and the Bossard-Nicolai (BN) model were discussed in
detail. An interesting insight into the construction of self-dual theories
was provided by Ivanov and Zupnik \cite{Ivanov:2003uj}, \cite{Ivanov:2012bq}
using the bispinor auxiliary fields. The authors also proved, that this
approach appears to be fully equivalent to the one based on the nonlinear
twisted self-duality constraint. Both the latter two approaches parametrize
the general solution of the NGZ condition in terms of one arbitrary
functional which has manifest $U(1)$ rotational symmetry. However, the
physical meaning of this functional is not completely clear. Note also that,
in spite of the progress in understanding the self-duality, only few
Lagrangians leading to self-dual theories beyond the BI one are known in a
closed form. For instance for the Lagrangian of the BN model only the first
eight terms of the weak field expansion has been calculated explicitly \cite%
{Carrasco:2011jv} and its closed form is not known yet.

The BI theory is a prominent member of the class of self-dual theories and
since its birth it has been subject of countless studies. The renewed
interest in this theory was inspired by strings and D-branes: as was shown
by Fradkin and Tseytlin in \cite{Fradkin:1985qd}, the BI Lagrangian can be
interpreted as a low energy effective theory describing the vector field
coupled to the string ending on a D-brane (see also \cite{Tseytlin:1996it}
and \cite{Tseytlin:1999dj}). It also naturally appears as a bosonic sector
of the effective theory which corresponds to the spontaneous breaking of the 
${\mathcal{N}}=2$ to ${\mathcal{N}}=1$ SUSY (see \cite{Bagger:1996wp}, \cite%
{Rocek:1997hi} and \cite{Tseytlin:1999dj}). Also the tree level amplitudes
in BI theory are special. First, they conserve helicity, i.e. the amplitudes
with non-equal number of helicity plus and helicity minus photons (when all
particles are assumed to be outgoing) vanish identicaly. This was proved in 
\cite{Rosly:2002jt}, where this property was interpreted as a consequence of
the self-duality of the theory, and independently in \cite{Boels:2008fc}
using Feynman diagrams. This results allow to conclude, that the helicity
violating one loop amplitudes in the BI theory have vanishing imaginary
parts and should be rational functions which hopefully vanish too. However,
the general proof of the possible helicity conservation of the higher loop
amplitudes in BI theory still seems to be an open problem. The second
interesting property of tree level amplitudes in BI theory has been
established quite recently in \cite{Cheung:2018oki}. Namely, the tree level
amplitude have an unique soft behavior (they vanish) in the multi-chiral
soft limit when all the particles with the same helicity\footnote{%
Here we again assume all the particles to be outgoing.} become
simultaneously soft. This soft behavior constraints the amplitude strongly
enough to fix uniquely the BI theory (up to the choice of the units\footnote{%
I.e. up to one dimensionful parameter which corresponds at the classical
level to the maximal intensity of the electric field.}). Note also, that the
BI belongs to the class of theories tree amplitudes of which admit the CHY
representation \cite{Cachazo:2014xea}.

It is a natural question which of the above properties of the BI theory are
connected intimately with the self-duality only and can be thus proved for
any self-dual theory. In this study we concentrate mainly on the connection
of the helicity conservation and self-duality at the tree and loop level in
the general nonlinear QED. As a byproduct we find a physical interpretation
of the $U(1)$ rotational invariant generating functional which appears in
the auxiliary field method of Ivanov and Zupnik (or equivalently in the
method of nonlinear twisted self-duality constraint of Carrasco, Kallosh and
Roiban) in terms of a certain generalization of normal ordering, which
simplifies the Feynman rules for perturbative calculation of the S-matrix.
As an explicit example we find a normal ordered version of the BI Lagrangian
in a closed form (and re-derive at the same time the hypergeometric form of
the BI Lagrangian found originally in \cite{Aschieri:2013nda} and \cite%
{Aschieri:2013aza}) and calculate a semi-closed form of the BN Lagrangian
(in terms of infinite series corresponding to the weak field expansion with
explicit coefficients). We also briefly discuss a general form of the
self-dual Lagrangian in terms of its normal ordered form.

The article is organized as follows. In section 2 we shortly remind the
basics of the nonlinear QED and duality transformation and also fix our
notation. In section 3 we briefly discuss various representation of the general
solutions of the NGZ identity (including a new one) and give some examples
of the self-dual Lagrangians beyond the Maxwell and BI case. In Section 4 we
discuss the quantization of nonlinear QED with stress on various versions of
the Feynman rules within different representations of the Lagrangian.
Section 5 is devoted to the proof of the tree-level helicity conservation in
a general self-dual QED. In Section 6 we introduce the normal ordering and
modified Feynman rules and discuss the helicity conservation at the loop
level. We also find the normal ordered form of the BI Lagrangian and the
semi-closed form of the BN Lagrangian and give a general prescription for
the transformation of the normal ordered Lagrangian into the usual form. In
Section 7 we summarize the results.

\section{Nonlinear electrodynamics and duality}

In what follows we will consider models of the nonlinear electrodynamic in
four dimensions the Lagrangian of which is a functions of the field strength
tensor $F_{\mu \nu }=\partial _{\mu }A_{\nu }-\partial _{\nu }A_{\mu }$
only. The most general such Lagrangian can be written in the form%
\begin{equation}
\mathcal{L}=-\frac{1}{4}F_{\mu \nu }F^{\mu \nu }+\mathcal{L}_{int}\left(
F_{\mu \nu }\right) ,  \label{general_model}
\end{equation}%
where $\mathcal{L}_{int}\left( F_{\mu \nu }\right) =O\left( F^{4}\right) $.
From the phenomenological point of view, such models can appear as the
leading order in the derivative expansion of the nonlocal effective action
obtained by means of integrating out the massive charged degrees of freedom.
Let us mention in this context the famous Euler-Heisenberg Lagrangian \cite%
{Heisenberg:1935qt} (see also \cite{Dunne:2004nc} for a comprehensive
review) which describes effective interactions of the low-energy photons at
energy scale $p\ll m_{e}$ where $m_{e}$ is the electron mass,%
\begin{equation}
\mathcal{L}_{int}^{EH}\left( F_{\mu \nu }\right) =\frac{\alpha ^{2}}{%
90m_{e}^{4}}\left[ \left( F^{\mu \nu }F_{\mu \nu }\right) ^{2}+\frac{7}{4}%
\left( \widetilde{F}^{\mu \nu }F_{\mu \nu }\right) ^{2}\right] +O\left(
\left( \frac{\alpha }{m_{e}}\right) ^{3}\right) .
\end{equation}%
Here%
\begin{equation}
\widetilde{F}^{\mu \nu }=\frac{1}{2}\varepsilon ^{\mu \nu \alpha \beta
}F_{\alpha \beta }
\end{equation}%
is the dual field strengths and we have written explicitly only the leading
term in the fine structure constant $\alpha $. Another example is the
Born-Infeld modification of the Maxwell electrodynamics \cite{Born:1934gh}
designed originally in order to solve the problem of the infinite
electromagnetic self-energy of the point charge%
\begin{eqnarray}
\mathcal{L}^{BI} &=&-\Lambda ^{4}\sqrt{-\det \left( \eta _{\mu \nu }+\frac{%
F_{\mu \nu }}{\Lambda ^{2}}\right) }+\Lambda ^{4}  \notag \\
&=&-\Lambda ^{4}\sqrt{1+\frac{2}{\Lambda ^{4}}\mathcal{F}-\frac{1}{\Lambda
^{8}}\mathcal{G}^{2}}+\Lambda ^{4}.
\end{eqnarray}%
Here the two independent invariants $\mathcal{F}$ and $\mathcal{G}$ (in four
dimensions any other invariant is a function of these two) are defined as%
\begin{equation}
\mathcal{F}=\frac{1}{4}F_{\mu \nu }F^{\mu \nu },~~~~\mathcal{G}=\frac{1}{4}%
F_{\mu \nu }\widetilde{F}^{\mu \nu }.
\end{equation}%
The dimensionful scale $\Lambda $ sets a limit on the maximal possible
intensity of the electric field, $E_{\max }=\Lambda ^{2}$. The Lagrangian $%
\mathcal{L}^{BI}$ also appears as an effective action disribing fluctuations
of the massless degrees of freedom of an open string ending on a D-brane
corresponding to the string excitations longitudinal to the brane. In this
context $\Lambda ^{-2}=2\pi \alpha ^{\prime }=T^{-1}$ where $T$ is the
string tension and $\alpha ^{\prime }$ is the Regge slope.

For further consideration it will be useful to reformulate the Lagrangian in
terms of the symmetric spinor fields $\phi _{AB}$ and $\overline{\phi }_{%
\overset{.}{A}\overset{.}{B}}$ defined as%
\begin{equation}
F_{\mu \nu }\overline{\sigma }_{A\overset{.}{A}}^{\mu }\overline{\sigma }_{B%
\overset{.}{B}}^{\nu }=\phi _{AB}\varepsilon _{\overset{.}{A}\overset{.}{B}}+%
\overline{\phi }_{\overset{.}{A}\overset{.}{B}}\varepsilon _{AB}
\label{spinor_definition}
\end{equation}%
where as usual $\sigma ^{\mu }=\left( 1,\boldsymbol{\sigma }\right) $ and $%
\overline{\sigma }^{\mu }=\left( 1,-\boldsymbol{\sigma }\right) $. Let us
note that%
\begin{equation}
\widetilde{F}_{\mu \nu }\overline{\sigma }_{A\overset{.}{A}}^{\mu }\overline{%
\sigma }_{B\overset{.}{B}}^{\nu }=\mathrm{i}\phi _{AB}\varepsilon _{\overset{%
.}{A}\overset{.}{B}}-\mathrm{i}\overline{\phi }_{\overset{.}{A}\overset{.}{B}%
}\varepsilon _{AB}
\end{equation}%
As a consequence of the Cayley-Hamilton theorem for two-by-two traceless
matrices $\boldsymbol{\Phi }\equiv \phi _{~~B}^{A}=\varepsilon ^{AC}\phi
_{CB}$ and $\overline{\boldsymbol{\Phi }}\equiv \overline{\phi }_{~~\overset{%
.}{B}}^{\overset{.}{A}}=\varepsilon ^{\overset{.}{A}\overset{.}{C}}\overline{%
\phi }_{\overset{.}{C}\overset{.}{B}}$ we get%
\begin{equation}
\boldsymbol{\Phi }^{2} =-\det \boldsymbol{\Phi =}-\frac{1}{2}\phi ^{AD}\phi
_{AD}\equiv -\frac{1}{2}\phi ^{2}
\end{equation}%
and thus%
\begin{equation}
\mathrm{Tr}\boldsymbol{\Phi }^{2n}=\mathrm{Tr}\left( -\frac{1}{2}\phi
^{2}\right) ^{n}=\frac{\left( -1\right) ^{n}}{2^{n-1}}\left( \phi
^{2}\right) ^{n},~~~~~\mathrm{Tr}\boldsymbol{\Phi }^{2n+1}=0
\end{equation}%
and similarly for $\overline{\boldsymbol{\Phi }}$. Therefore the most
general invariant built from $F_{\mu \nu }$ only can be expressed as a
function of two independent invariants $\phi ^{2}$ and $\overline{\phi }^{2}$%
. For instance for the above two invariants $\mathcal{F}$ and $\mathcal{G}$
we get%
\begin{equation}
\mathcal{F=}\frac{1}{8}\left( \phi ^{2}+\overline{\phi }^{2}\right) ,~~~~~%
\mathcal{G}=\frac{\mathrm{i}}{8}\left( \phi ^{2}-\overline{\phi }^{2}\right)
,  \label{F_G_in_terms_of_phi_phibar}
\end{equation}%
and the most general Lagrangian (\ref{general_model}) can be expressed as a
function of two variables $\mathcal{L}_{int}\left( \phi ^{2},\overline{\phi }%
^{2}\right) $ in the form 
\begin{equation}
\mathcal{L}=-\frac{1}{8}\left( \phi ^{2}+\overline{\phi }^{2}\right) +%
\mathcal{L}_{int}\left( \phi ^{2},\overline{\phi }^{2}\right) ,
\label{L_in_phi_phibar}
\end{equation}%
where%
\begin{equation}
\mathcal{L}_{int}=\sum_{n+m>1}c_{nm}\left( \phi ^{2}\right) ^{n}\left( 
\overline{\phi }^{2}\right) ^{m}  \label{general_interaction_phi_phibar}
\end{equation}%
and where hermiticity requires $c_{nm}^{\ast }=c_{mn}$.

The classical equations of motion without sources expressed in terms of $%
F_{\mu \nu }$ consist of the Bianchi identity and the Euler-Lagrange equation%
\begin{equation}
\partial _{\mu }\widetilde{F}^{\mu \nu }=0,~\ ~~~~\partial _{\mu }G^{\mu \nu
}=0  \label{EOM}
\end{equation}%
where%
\begin{equation}
G_{\mu \nu }=-2\frac{\partial \mathcal{L}}{\partial F^{\mu \nu }}
\label{G_definition}
\end{equation}%
is the Lorentz covariant constitutive equation. The above equations (\ref%
{EOM}) are invariant with respect to the famous duality transformation
written in the infinitesimal form as%
\begin{eqnarray}
\delta F_{\mu \nu } &=&\widetilde{G}_{\mu \nu }=\frac{1}{2}\varepsilon _{\mu
\nu \alpha \beta }G^{\alpha \beta },  \label{deltaF} \\
\delta G_{\mu \nu } &=&\widetilde{F}_{\mu \nu }=\frac{1}{2}\varepsilon _{\mu
\nu \alpha \beta }F^{\alpha \beta },  \label{deltaG}
\end{eqnarray}%
provided the Lagrangian satisfies the Noether-Gaillard-Zumino (NGZ) relation 
\cite{Gaillard:1981rj}%
\begin{equation}
\varepsilon ^{\mu \nu \alpha \beta }\frac{\partial \mathcal{L}}{\partial
F^{\mu \nu }}\frac{\partial \mathcal{L}}{\partial F^{\alpha \beta }}=\frac{1%
}{4}\varepsilon _{\mu \nu \alpha \beta }F^{\mu \nu }F^{\alpha \beta }+C,
\label{NGZ_relation_F}
\end{equation}%
where $C$ is an arbitrary constant. The latter relation expresses
consistency of the transformation (\ref{deltaF}) and (\ref{deltaG}) with
definition of $G_{\mu \nu }~$(\ref{G_definition}). Provided we require the
weak field expansion of the constitutive relation of the form%
\begin{equation*}
G_{\mu \nu }=F_{\mu \nu }+O(F^{2}),
\end{equation*}%
i.e. the theory can be approximated \ in this limit by Maxwell
electromagnetism, we get for the constant $C=0$. In what follows we restrict
ourselves to this case and refer to the theories satisfying (\ref%
{NGZ_relation_F}) with $C=0$ as self-dual theories.

Note however, that the duality transformation (\ref{deltaF}), (\ref{deltaG})
is not an off-shell symmetry of the Lagrangian, but an on-shell symmetry of
the equations of motion.

In terms of the spinors $\phi _{AB}$ and $\overline{\phi }_{\overset{.}{A}%
\overset{.}{B}}$ and analogous spinors $\Gamma _{AB}$ and $\overline{\Gamma }%
_{\overset{.}{A}\overset{.}{B}}$where%
\begin{eqnarray}
\overline{\sigma }_{A\overset{.}{A}}^{\mu }\overline{\sigma }_{B\overset{.}{B%
}}^{\nu }G_{\mu \nu } &=&\Gamma _{AB}\varepsilon _{\overset{.}{A}\overset{.}{%
B}}+\overline{\Gamma }_{\overset{.}{A}\overset{.}{B}}\varepsilon _{AB} \\
\Gamma _{AB} &=&-8\frac{\partial \mathcal{L}}{\partial \phi ^{2}}\phi _{AB},
\\
\overline{\Gamma }_{\overset{.}{A}\overset{.}{B}} &=&-8\frac{\partial 
\mathcal{L}}{\partial \overline{\phi }^{2}}\overline{\phi }_{\overset{.}{A}%
\overset{.}{B}}
\end{eqnarray}%
we can rewrite the duality transformation as%
\begin{eqnarray}
\delta \phi _{AB} &=&-\mathrm{i}\Gamma _{AB},~~~~\delta \overline{\phi }_{%
\overset{.}{A}\overset{.}{B}}=\mathrm{i}\overline{\Gamma }_{\overset{.}{A}%
\overset{.}{B}}, \\
\delta \Gamma _{AB} &=&\mathrm{i}\phi _{AB},~~~~\delta \overline{\Gamma }_{%
\overset{.}{A}\overset{.}{B}}=-\mathrm{i}\overline{\phi }_{\overset{.}{A}%
\overset{.}{B}}.
\end{eqnarray}%
The NGZ relation (\ref{NGZ_relation_F}) in these variables reads (note that
the Lagrangian is a function of the invariants $\phi ^{2}$ and $\overline{%
\phi }^{2}$)%
\begin{equation}
\phi ^{2}\left( \frac{\partial \mathcal{L}}{\partial \phi ^{2}}\right) ^{2}-%
\overline{\phi }^{2}\left( \frac{\partial \mathcal{L}}{\partial \overline{%
\phi }^{2}}\right) ^{2}-\frac{1}{64}\left( \phi ^{2}-\overline{\phi }%
^{2}\right) =0.  \label{NGZ}
\end{equation}%
It is straightforward to verify that the Born-Infeld Lagrangian, which in
the variables $\phi _{AB}$ and $\overline{\phi }_{\overset{.}{A}\overset{.}{B%
}}$ reads 
\begin{equation}
\mathcal{L}_{BI}=-\Lambda ^{4}\sqrt{1+\frac{1}{4\Lambda ^{4}}\left( \phi
^{2}+\overline{\phi }^{2}\right) +\frac{1}{64\Lambda ^{8}}\left( \phi ^{2}-%
\overline{\phi }^{2}\right) ^{2}}+\Lambda ^{4},
\label{L_BI_original_phi_phibar}
\end{equation}%
satisfy the NGZ relation (\ref{NGZ}) and the theory is therefore self-dual.

The NGZ relation (\ref{NGZ}) can be further simplified using formally the
variables%
\begin{equation}
X_{\pm }=\frac{1}{2}\left( \sqrt{\phi ^{2}}\pm \sqrt{\overline{\phi }^{2}}%
\right)
\end{equation}%
to the form%
\begin{equation}
\frac{\partial \mathcal{L}}{\partial X_{+}}\frac{\partial \mathcal{L}}{%
\partial X_{-}}=\frac{1}{4}X_{+}X_{-}
\end{equation}%
or 
\begin{equation}
\frac{\partial \mathcal{L}}{\partial X_{+}^{2}}\frac{\partial \mathcal{L}}{%
\partial X_{-}^{2}}=\frac{1}{16}.  \label{duality_diff_equation}
\end{equation}%
This is the most suitable form for further consideration. In the next
section we will discuss the solution of this equation in more detail and
give some explicit examples of self-dual Lagrangians beyond the BI theory.

\section{General solutions of the NGZ identity}

The NGZ identity written in the form (\ref{duality_diff_equation}) is a
partial differential equation of the first order and as such it can be
solved using standard methods. Of course not all of its solutions are
physically acceptable. We typically require analyticity of the resulting
Lagrangian in the variables $\phi ^{2}$ and $\overline{\phi }^{2}$ at the
origin and we also expect that the weak field limit should reproduce the
Maxwell electrodynamics. In this section we give a general prescription and
also formulate the necessary condition for the above analyticity requirement.

According to the general methods for solution of the first order partial
differential equations by means of characteristics, the general solution $%
\mathcal{L}\left( X_{+},X_{-}\right) $ of the equation (\ref%
{duality_diff_equation}) can be expressed implicitly in terms of four
functions $p_{\pm }\left( u\right) $ and $x_{\pm }\left( u\right) $ which
play the role of the one parametric set of the initial values of the
characteristics, namely%
\begin{eqnarray}
\pm 4\mathcal{L}\left( X_{+},X_{-}\right) &=&p_{-}\left( u\right) \left[
X_{-}^{2}-x_{-}\left( u\right) \right] +p_{+}\left( u\right) \left[
X_{+}^{2}-x_{+}\left( u\right) \right]  \notag \\
&&+\int \mathrm{d}u\left[ p_{+}\left( u\right) x_{+}^{\prime }\left(
u\right) +p_{-}\left( u\right) x_{-}^{\prime }\left( u\right) \right] .
\end{eqnarray}%
Here the prime denote a derivative with respect to the parameter $u$. These
functions are subject of the constraints%
\begin{eqnarray}
p_{+}\left( u\right) p_{-}\left( u\right) &=&1  \label{1st_constraint} \\
p_{+}\left( u\right) \left[ X_{+}^{2}-x_{+}\left( u\right) \right]
&=&p_{-}\left( u\right) \left[ X_{-}^{2}-x_{-}\left( u\right) \right] .
\label{2nd_constraint}
\end{eqnarray}%
The first constraint reduces the number of independent function to three
while the second one allows to determine the parameter $u$ in terms of the
variables $X_{\pm }$. For instance, the Maxwell theory can be reproduced by
the choice%
\begin{equation}
x_{\pm }^{M}\left( u\right) =0,~~~p_{\pm }^{M}\left( u\right) =-1.
\end{equation}%
Because the functions $p_{\pm }\left( u\right) $ and $x_{\pm }\left(
u\right) $ appear in the above expressions in very special combinations, the
above formula can be further simplified in such a way that there is only one
arbitrary function left. Using integration by parts we get%
\begin{eqnarray}
\pm 4\mathcal{L}\left( X_{+},X_{-}\right) &=&p_{-}\left( u\right)
X_{-}^{2}+p_{+}\left( u\right) X_{+}^{2}  \notag \\
&&-\int \mathrm{d}u\left[ p_{+}^{\prime }\left( u\right) x_{+}\left(
u\right) +p_{-}^{\prime }\left( u\right) x_{-}\left( u\right) \right] .
\end{eqnarray}%
Writing the explicit solution of the first constraint (\ref{1st_constraint})
in the form%
\begin{equation}
p_{-}\left( u\right) =p\left( u\right) ,~~~~p_{+}\left( u\right) =\frac{1}{%
p\left( u\right) }
\end{equation}%
we have%
\begin{equation}
\pm 4\mathcal{L}\left( X_{+},X_{-}\right) =p\left( u\right) X_{-}^{2}+\frac{1%
}{p\left( u\right) }X_{+}^{2}+\int \mathrm{d}u\frac{p^{\prime }\left(
u\right) }{p\left( u\right) }F\left( u\right) .  \label{p(u)_representation}
\end{equation}%
where $u$ is the solution of the second constraint (\ref{2nd_constraint}),
which can be written in the form 
\begin{equation}
\frac{1}{p\left( u\right) }X_{+}^{2}-p\left( u\right) X_{-}^{2}=F\left(
u\right)  \label{p(u)_euqation}
\end{equation}%
and the function $F\left( u\right) $ is defined as%
\begin{equation}
F\left( u\right) \equiv \frac{1}{p\left( u\right) }x_{+}\left( u\right)
-p\left( u\right) x_{-}\left( u\right) .  \label{F(u)_definition}
\end{equation}%
Introducing a new variable $z=p\left( u\right) $ and denoting $f\left(
z\right) =F\left( u\left( z\right) \right) $ we can rewrite the second
constraint (\ref{2nd_constraint})\ as%
\begin{equation}
\frac{1}{z}X_{+}^{2}-zX_{-}^{2}=f\left( z\right)  \label{z_equation}
\end{equation}%
and finally we get the Lagrangian represented implicitly in terms of one
arbitrary function $f\left( z\right) $ 
\begin{equation}
\pm 4\mathcal{L}\left( X_{+},X_{-}\right) =zX_{-}^{2}+\frac{1}{z}%
X_{+}^{2}+\int \frac{\mathrm{d}z}{z}f\left( z\right) ,  \label{z_Lagrangian}
\end{equation}%
where $z$ is the solution of (\ref{z_equation}).

In what follows we will mainly restrict ourselves to the case when $\mathcal{%
L}\left( X_{+},X_{-}\right) $ is analytic in $\phi ^{2}$ and $\overline{\phi 
}^{2}$. Because the resulting Lagrangian (\ref{z_Lagrangian}) depends on $%
X_{\pm }$ only through\footnote{%
Note also that $X_{\pm }^{2}=2\left( {\mathcal{F}}\pm \sqrt{{\mathcal{F}}%
^{2}+{\mathcal{G}}^{2}}\right) $.} $X_{\pm }^{2}$, assuming analyticity in $%
X_{\pm }^{2}$ the necessary condition for such an analyticity can be
expressed as a symmetry condition%
\begin{equation}
\mathcal{L}\left( X_{+},X_{-}\right) =\mathcal{L}\left( X_{-},X_{+}\right) .
\end{equation}%
This can be achieved by the choice of function $f\left( z\right) $ which
satisfies%
\begin{equation}
f\left( z\right) =-f\left( \frac{1}{z}\right) .  \label{f_symmetry}
\end{equation}%
Indeed, in such a case the solution of (\ref{z_equation}) with $X_{+}$ and $%
X_{-}$ interchanged is just $1/z$ where $z$ is the original solution of (\ref%
{z_equation}) and the sum of the first two terms on the right hand side of (%
\ref{z_Lagrangian}) is therefore invariant. For the second term we get
immediately, provided we fix the lower limit of the integration appropriately%
\footnote{%
Here we tacitly assume that either $f\left( 1\right) =0$ or $f\left(
z\right) $ has at most integrable singularity for $z=1$}, 
\begin{equation}
\int_{1}^{1/z}\frac{\mathrm{d}u}{u}f\left( u\right) =-\int_{1}^{z}\frac{%
\mathrm{d}w}{w^{2}}wf\left( \frac{1}{w}\right) =\int_{1}^{z}\frac{\mathrm{d}w%
}{w}f\left( w\right) .
\end{equation}

Let us now give some simple examples. Taking 
\begin{equation}
f^{BI}\left( z\right) =2\Lambda ^{4}\left( z-\frac{1}{z}\right)
\end{equation}%
and arranging the integration constant we reproduce the BI Lagrangian%
\begin{equation}
\mathcal{L}^{BI}=\mathcal{-}\Lambda ^{4}\sqrt{1+\frac{1}{2\Lambda ^{4}}%
\left( X_{+}^{2}+X_{-}^{2}\right) +\frac{1}{4\Lambda ^{8}}X_{+}^{2}X_{-}^{2}}%
+\Lambda ^{4}.
\end{equation}%
The apparently simplest one parametric deformation of the BI Lagrangian can
be obtained in this representation using%
\begin{equation*}
f^{MBI}\left( z,a\right) =f^{BI}\left( z\right) -4a\Lambda ^{4}.
\end{equation*}%
The resulting Lagrangian reads%
\begin{equation}
\mathcal{L}^{MBI}=-\Lambda ^{4}\left[ r\left( X_{+}^{2},X_{-}^{2}\right)
-a\ln \left( \frac{a+r\left( X_{+}^{2},X_{-}^{2}\right) }{1+\frac{1}{%
2\Lambda ^{4}}X_{-}^{2}}\right) -c\right]
\end{equation}%
where%
\begin{eqnarray}
r\left( X_{+}^{2},X_{-}^{2}\right) &=&\sqrt{1+a^{2}+\frac{1}{2\Lambda ^{4}}%
\left( X_{+}^{2}+X_{-}^{2}\right) +\frac{1}{4\Lambda ^{8}}X_{+}^{2}X_{-}^{2}}%
, \\
c &=&\sqrt{1+a^{2}}-a\ln \left( a+\sqrt{1+a^{2}}\right) .
\end{eqnarray}%
Note however, that because $f^{MBI}\left( z,a\right) $ does not satisfy (\ref%
{f_symmetry}) the Lagrangian $\mathcal{L}^{MBI}$ is not analytic for $\phi
^{2}=\overline{\phi }^{2}=0$. Indeed, the weak field expansion reads now%
\begin{eqnarray}
\mathcal{L}^{MBI} &=&-\frac{\sqrt{1+a^{2}}}{8}\left( \phi ^{2}+\overline{%
\phi }^{2}\right) +\frac{a}{4}\sqrt{\phi ^{2}\overline{\phi }^{2}}-\frac{a}{%
32\Lambda ^{4}}\sqrt{\phi ^{2}\overline{\phi }^{2}}\left( \phi ^{2}+%
\overline{\phi }^{2}\right)  \notag \\
&&+\frac{a^{2}}{128\Lambda ^{4}\sqrt{1+a^{2}}}\left[ \phi ^{4}+\overline{%
\phi }^{4}+2\left( 3-2a^{-2}\right) \phi ^{2}\overline{\phi }^{2}\right]
+\ldots
\end{eqnarray}%
Note also the non-canonical normalization of the kinetic term.

Let us now relate the above construction of the self-dual Lagrangian to
those known from the literature. The representation (\ref{z_Lagrangian}), (%
\ref{z_equation}) can be compared with the general solution found in \cite%
{Gaillard:1997rt} defining a new variable $w=-f\left( z\right) /z$.
Expressing $z$ in terms of $w$ we get%
\begin{equation}
\pm 4\mathcal{L}\left( X_{+},X_{-}\right) =z\left( w\right) X_{-}^{2}+\frac{1%
}{z\left( w\right) }X_{+}^{2}-wz\left( w\right) +\int \mathrm{d}wz\left(
w\right)  \label{w_representation}
\end{equation}%
and $w$ is a solution of 
\begin{equation}
X_{-}^{2}=\frac{1}{z\left( w\right) ^{2}}X_{+}^{2}+w.
\end{equation}%
Finally, using this equation in (\ref{w_representation}) we get%
\begin{equation}
\pm 4\mathcal{L}\left( X_{+},X_{-}\right) =\frac{2}{z\left( w\right) }%
X_{+}^{2}+\int \mathrm{d}wz\left( w\right)
\end{equation}%
which is nothing else but the Gaillard-Zumino representation which expresses
the solution in terms of arbitrary function $z\left( w\right) $. The latter
representation has the advantage that it allows to find the function $%
z\left( w\right) $ once the Lagrangian $\mathcal{L}\left( X_{+},X_{-}\right) 
$ is known: for $X_{+}=0$ we get $w=X_{-}^{2}$ and thus%
\begin{equation}
z\left( w\right) =\pm \frac{\mathrm{d}}{\mathrm{d}w}4\mathcal{L}\left( 0,%
\sqrt{w}\right) .
\end{equation}%
For instance, for the BI theory we get immediately 
\begin{equation}
z^{BI}\left( w\right) =\frac{1}{\sqrt{1+\frac{1}{2\Lambda ^{4}}w}}.
\end{equation}

Let us now define in (\ref{z_equation}) and (\ref{z_Lagrangian}) the
following variable%
\begin{equation}
u=\frac{z^{2}-1}{z^{2}+1}
\end{equation}%
and define in terms of this variable the following new function%
\begin{equation}
G\left( u\right) =\frac{1}{2}\frac{zf\left( z\right) }{z^{2}+1}.
\end{equation}%
It is then an easy exercise to rewrite (\ref{z_equation}) into the form%
\begin{equation}
\frac{1}{4}\left( X_{+}^{2}-X_{-}^{2}\right) -\frac{1}{4}\left(
X_{+}^{2}+X_{-}^{2}\right) u=G\left( u\right)  \label{kanimura_equation}
\end{equation}%
and (\ref{z_Lagrangian}) 
\begin{eqnarray}
\pm \mathcal{L}\left( X_{+},X_{-}\right) &=&\frac{1}{4}X_{+}^{2}\left( \sqrt{%
1-u^{2}}-\frac{u\left( 1-u\right) }{\sqrt{1-u^{2}}}\right) +\frac{1}{4}%
X_{-}^{2}\left( \sqrt{1-u^{2}}+\frac{u\left( 1+u\right) }{\sqrt{1-u^{2}}}%
\right)  \notag \\
&&+\int \mathrm{d}u\frac{G\left( u\right) }{\left( 1-u^{2}\right) ^{3/2}}.
\end{eqnarray}%
Using now (\ref{kanimura_equation}) we get%
\begin{equation}
\pm \mathcal{L}\left( X_{+},X_{-}\right) =\frac{1}{4}\left(
X_{+}^{2}+X_{-}^{2}\right) \sqrt{1-u^{2}}-\frac{uG\left( u\right) }{\sqrt{%
1-u^{2}}}+\int \mathrm{d}u\frac{G\left( u\right) }{\left( 1-u^{2}\right)
^{3/2}}
\end{equation}%
and finally%
\begin{equation}
\pm \mathcal{L}\left( X_{+},X_{-}\right) =\frac{1}{4}\left(
X_{+}^{2}+X_{-}^{2}\right) \sqrt{1-u^{2}}-\int \mathrm{d}u\frac{uG^{\prime
}\left( u\right) }{\sqrt{1-u^{2}}}.  \label{kanimura_lagrangian}
\end{equation}%
The latter formula together with the algebraic equation (\ref%
{kanimura_equation}) corresponds to Hatsuda-Kamimura-Sekiya representation
of the self-dual Lagrangian developed in \cite{Hatsuda:1999ys} in terms of
arbitrary function $G\left( u\right) $. Note that under the transformation $%
z\rightarrow 1/z$ the variable $u$ transforms as $u\rightarrow -u$ . Thus we
get for the function $G\left( u\right) $%
\begin{equation}
G\left( -u\right) =\frac{1}{2}\frac{zf\left( \frac{1}{z}\right) }{z^{2}+1}
\end{equation}%
and the necessary condition for analyticity of the Lagrangian reads now%
\begin{equation}
G\left( -u\right) =-G\left( u\right) .  \label{odd_G}
\end{equation}%
The BI Lagrangian is then reconstructed using%
\begin{equation*}
G^{BI}\left( u\right) =\Lambda ^{4}u
\end{equation*}
and in \cite{Hatsuda:1999ys} four more explicit examples were given, Note
however, that only one of them (namely the example 4 with $%
G(u)=u(1+au^2/3)/b $) satisfied the condition (\ref{odd_G}) and lead to the
analytic Lagrangian.

\section{Quantization of the nonlinear QED}

The usual formulation of the perturbation theory for the nonlinear
electrodynamics at the quantum level requires a gauge fixing. This procedure
sets the form of the propagator which then corresponds to the internal lines
of the Feynman graphs. In the Feynman gauge, which is manifestly Lorentz
covariant, we get a simple propagator of the form%
\begin{equation}
\langle 0|TA^{\mu }\left( x\right) A^{\nu }\left( y\right) |0\rangle =%
\mathrm{i}\int \frac{\mathrm{d}^{4}p}{\left( 2\pi \right) ^{4}}\mathrm{e}^{-%
\mathrm{i}p\cdot \left( x-y\right) }\frac{\eta ^{\mu \nu }}{p^{2}+\mathrm{i}0%
}  \label{AA_propagator}
\end{equation}%
The Feynman rules for the vertices are read off from the interaction
Lagrangian $\mathcal{L}_{int}\left( F_{\mu \nu }\right) $ treated as a
functional of the field $A_{\mu }\left( x\right) $ and the polarization
vectors $\varepsilon _{h}^{\mu }\left( p\right) $ and their complex
conjugates are attached to the incoming and outcoming external on-shell
lines respectively. However, for practical purposes of amplitude
calculation, the direct manipulation with the field $A^{\mu }\left( x\right) 
$ is rather clumsy because for the simple form of the propagator we have to
pay with relatively complicated (usually infinitely many) interaction
vertices which depend on the derivatives of $A^{\mu }\left( x\right) $. For
the general interaction Lagrangian of the form $\mathcal{L}_{int}\left(
F_{\mu \nu }\right) $ it is therefore much more convenient to work directly
with the field $F^{\mu \nu }\left( x\right) $. The covariant propagator of
the field $F^{\mu \nu }\left( x\right) $ can be derived from (\ref%
{AA_propagator}) by means of taking appropriate derivatives with the result 
\begin{equation}
\langle 0|TF^{\mu \nu }\left( x\right) F^{\alpha \beta }\left( y\right)
|0\rangle =\mathrm{i}\int \frac{\mathrm{d}^{4}p}{\left( 2\pi \right) ^{4}}%
\mathrm{e}^{-\mathrm{i}p\cdot \left( x-y\right) }\frac{P^{\mu \nu \alpha
\beta }\left( p\right) }{p^{2}+\mathrm{i}0},  \label{F_propagator}
\end{equation}%
where $P^{\mu \nu \alpha \beta }\left( p\right) $ is given by the expression 
\begin{equation}
P^{\mu \nu \alpha \beta }\left( p\right) =-p^{\mu }p^{\alpha }\eta ^{\nu
\beta }+p^{\mu }p^{\beta }\eta ^{\nu \alpha }+p^{\nu }p^{\alpha }\eta ^{\mu
\beta }-p^{\nu }p^{\beta }\eta ^{\mu \alpha }.
\end{equation}%
Note however, that the propagator of the field $F^{\mu \nu }\left( x\right) $
cannot be derived from any local kinetic term for the field $F_{\mu \nu
}\left( x\right) $. Indeed, the tensor $P^{\mu \nu \alpha \beta }\left(
p\right) $ can be rewritten as%
\begin{equation}
P^{\mu \nu \alpha \beta }\left( p\right) =p^{2}\left[ \frac{1}{2}\left( \eta
_{\mu \alpha }\eta _{\nu \beta }-\eta _{\mu \beta }\eta _{\nu \alpha
}\right) -\Pi _{\mu \nu \alpha \beta }^{T}\right] ,  \notag
\end{equation}%
where%
\begin{equation}
\Pi _{\mu \nu \alpha \beta }^{T}=\frac{1}{2}\left( P_{\mu \alpha }^{T}P_{\nu
\beta }^{T}-P_{\mu \beta }^{T}P_{\nu \alpha }^{T}\right) ,~~~~P_{\mu \nu
}^{T}=\eta _{\mu \nu }-\frac{p_{\mu }p_{\nu }}{p^{2}}.
\end{equation}%
Therefore $\Pi _{\mu \nu \alpha \beta }^{T}$ is the transverse projector in
the space of the antisymmetric tensors and $P^{\mu \nu \alpha \beta }\left(
p\right) $ is proportional to the longitudinal projector which has no
inversion.

Working directly with the fields $F_{\mu \nu }\left( x\right) $ the
interaction vertices are considerably simpler - they correspond to
non-derivative couplings of the field $F^{\mu \nu }\left( x\right) $, now
for the price of slightly more complicated propagator. Also the external
legs are now equipped with more complicated polarization tensors%
\begin{equation}
\varepsilon _{h}^{\mu \nu }\left( p\right) =-\mathrm{i}p^{\mu }\varepsilon
_{h}^{\nu }\left( p\right) +\mathrm{i}p^{\nu }\varepsilon _{h}^{\mu }\left(
p\right) .
\end{equation}%
Nevertheless the resulting Feynman rules for the $S-$matrix are completely
equivalent to those based on the propagator (\ref{AA_propagator}) and we get
manifest gauge invariance term by term for each Feynman diagram separately.

Even much more efficient treatment, which shares the latter property, is to
decompose the propagator of the field $F^{\mu \nu }\left( x\right) $ into
the spinor basis $\phi _{AB}\left( x\right) $ and $\overline{\phi }_{\overset%
{.}{C}\overset{.}{D}}\left( x\right) $ (cf.(\ref{spinor_definition})) The
free operators $\phi _{AB}\left( x\right) $ and $\overline{\phi }_{\overset{.%
}{C}\overset{.}{D}}\left( x\right) $ are directly connected with helicity: $%
\phi _{AB}\left( x\right) $ annihilates helicity plus and creates helicity
minus states while $\overline{\phi }_{\overset{.}{C}\overset{.}{D}}\left(
x\right) $ annihilates helicity minus and creates helicity plus states.
Other advantage of this decomposition is that the most general interaction
Lagrangian $\mathcal{L}_{int}\left( F_{\mu \nu }\right) $ can be rewritten
in the form (\ref{general_interaction_phi_phibar}) and therefore can be
treated as a function of $\phi ^{2}$ and $\overline{\phi }^{2}$. The
contraction of the spinor indices is thus considerably simpler than the
contraction of the original Lorentz indices and the structure of the
interaction vertices is then much more transparent within this formalism.

The decomposition of the propagator (\ref{F_propagator}) in the spinor basis
reads (see also \cite{Boels:2008fc}) 
\begin{eqnarray}
\langle 0|T\phi _{AB}\left( x\right) \overline{\phi }_{\overset{.}{C}\overset%
{.}{D}}\left( y\right) |0\rangle &=&\mathrm{i}\int \frac{\mathrm{d}^{4}p}{%
\left( 2\pi \right) ^{4}}\mathrm{e}^{-\mathrm{i}p\cdot \left( x-y\right) }%
\frac{\left[ p_{A\overset{.}{C}}p_{B\overset{.}{D}}+p_{A\overset{.}{D}}p_{B%
\overset{.}{C}}\right] }{p^{2}+\mathrm{i}0}  \label{propagator_mixed} \\
0|T\phi _{AB}\left( x\right) \phi _{CD}\left( y\right) |0\rangle &=&-\mathrm{%
i}\int \frac{\mathrm{d}^{4}p}{\left( 2\pi \right) ^{4}}\mathrm{e}^{-\mathrm{i%
}p\cdot \left( x-y\right) }\left[ \varepsilon _{AC}\varepsilon
_{BD}+\varepsilon _{AD}\varepsilon _{BC}\right]  \label{propagator_phi} \\
0|T\phi _{\overset{.}{A}\overset{.}{B}}\left( x\right) \phi _{\overset{.}{C}%
\overset{.}{D}}\left( y\right) |0\rangle &=&-\mathrm{i}\int \frac{\mathrm{d}%
^{4}p}{\left( 2\pi \right) ^{4}}\mathrm{e}^{-\mathrm{i}p\cdot \left(
x-y\right) }\left[ \varepsilon _{\overset{.}{A}\overset{.}{C}}\varepsilon _{%
\overset{.}{B}\overset{.}{D}}+\varepsilon _{\overset{.}{A}\overset{.}{D}%
}\varepsilon _{\overset{.}{B}\overset{.}{C}}\right].
\label{propagator_phi_bar}
\end{eqnarray}%
Here as usual%
\begin{equation}
p_{A\overset{.}{B}}=\overline{\sigma }_{A\overset{.}{B}}^{\mu }p_{\mu }
\end{equation}%
and $p_{A\overset{.}{B}}=p_{A}p_{\overset{.}{B}}$ on shell.

Note that only the \textquotedblleft mixed\textquotedblleft\ propagator $%
\langle T\phi \overline{\phi }\rangle $ posses a one-particle pole with
residue%
\begin{equation}
\left[ p_{A\overset{.}{C}}p_{B\overset{.}{D}}+p_{A\overset{.}{D}}p_{B\overset%
{.}{C}}\right] _{on-shell}=2p_{A}p_{B}p_{\overset{.}{A}}p_{\overset{.}{B}}.
\end{equation}%
The corresponding Feynman rules associate therefore the combination $\sqrt{2}%
p_{A}p_{B}$ to each outgoing helicity minus and ingoing helicity plus
external leg and $\sqrt{2}p_{\overset{.}{A}}p_{\overset{.}{B}}$ to each
outgoing helicity plus and ingoing helicity minus external leg. The
remaining two propagators (\ref{propagator_phi}) and (\ref%
{propagator_phi_bar}) are pure contact terms proportional to the delta
function of its space-time arguments\footnote{%
In fact these helicity violating propagators are unnecessary and can be
discarded by means of the procedure of normal ordering which we describe in
the next section.}. Their presence is a direct consequence of the simple
form of the original covariant propagators (\ref{AA_propagator}) and (\ref%
{F_propagator}) we have started with.

\section{Helicity conservation in nonlinear QED at tree level}

Let us now assume the $S-$ matrix of the general nonlinear QED. For our
purposes it is convenient to treat it as a functional $\mathcal{S}\left[
\phi ,\overline{\phi }\right] $ of the classical off-shell fields $\phi
_{AB}\left( x\right) $ and $\overline{\phi }_{\overset{.}{C}\overset{.}{D}%
}\left( x\right) $. The functional $\mathcal{S}\left[ \phi ,\overline{\phi }%
\right] $ is sometimes called the normal symbol of the $S-$ matrix: once we
know $\mathcal{S}\left[ \phi ,\overline{\phi }\right] $, the operator $S-$
matrix $\widehat{S}$ can be obtained by means of replacing the functional
arguments with free fields operators $\phi _{AB}\left( x\right) _{I}$ and $%
\overline{\phi }_{\overset{.}{C}\overset{.}{D}}\left( x\right) _{I}$ in the
interaction picture and then applying the usual normal ordering%
\begin{equation}
\widehat{S}=\colon \mathcal{S}\left[ \phi _{I},\overline{\phi }_{I}\right]
\colon
\end{equation}%
The analogous normal symbol $\mathcal{T}\left[ \phi ,\overline{\phi }\right] 
$ of the connected $S-$ matrix is related to $\mathcal{S}\left[ \phi ,%
\overline{\phi }\right] $ via the relation%
\begin{equation}
\mathcal{S}\left[ \phi ,\overline{\phi }\right] =\exp \left( \mathrm{i}%
\mathcal{T}\left[ \phi ,\overline{\phi }\right] \right)  \label{T_matrix}
\end{equation}%
The scattering amplitudes can be obtained directly form $\mathcal{T}\left[
\phi ,\overline{\phi }\right] $ applying appropriate differential operators
(for pedagogical treatment of this formalism see e.g. \cite{Siegel:1999ew}).

In what follows we will concentrate on theories the scattering amplitudes of
which conserve helicity. This means that, when we treat all the external
particles as outgoing, the amplitudes vanish provided the total number of
helicity plus particles does not match the total number of helicity minus
particles: 
\begin{equation}
A\left( 1^{+},2^{+},\ldots ,n^{+},(n+1)^{-},(n+2)^{-},\ldots
,(n+m)^{-}\right) =0~~~~\mathrm{for}~~~n\neq m.
\end{equation}%
Because the fields $\phi _{AB}\left( x\right) $ and $\overline{\phi }_{%
\overset{.}{C}\overset{.}{D}}\left( x\right) $ are associated with helicity
minus and helicity plus outgoing particles respectively, the requirement of
helicity conservation necessitates the functional $\mathcal{T}\left[ \phi ,%
\overline{\phi }\right] $ to be invariant with respect to the global $%
U\left( 1\right) $ transformation%
\begin{equation}
\phi _{AB}^{\prime }=\mathrm{e}^{\mathrm{i}\alpha }\phi _{AB},~~~~\overline{%
\phi }_{\overset{.}{A}\overset{.}{B}}^{\prime }=\mathrm{e}^{-\mathrm{i}%
\alpha }\overline{\phi }_{\overset{.}{A}\overset{.}{B}},
\end{equation}%
Infinitesimally this means%
\begin{equation}
\int \mathrm{d}^{4}x\left[ \phi _{AB}\left( x\right) \frac{\delta \mathcal{T}%
\left[ \phi ,\overline{\phi }\right] }{\delta \phi _{AB}\left( x\right) }-%
\overline{\phi }_{\overset{.}{A}\overset{.}{B}}\left( x\right) \frac{\delta 
\mathcal{T}\left[ \phi ,\overline{\phi }\right] }{\delta \overline{\phi }_{%
\overset{.}{A}\overset{.}{B}}\left( x\right) }\right] =0.
\label{U(1)_symmetry}
\end{equation}%
Let us now prove that duality invariance is necessary and sufficient
condition for helicity conservation of tree level amplitudes, i.e. that the
leading order term of the functional $\mathcal{T}\left[ \phi ,\overline{\phi 
}\right] $ in the quasiclassical expansion%
\begin{equation}
\mathcal{T}\left[ \phi ,\overline{\phi }\right] =\mathcal{T}^{tree}\left[
\phi ,\overline{\phi }\right] +O\left( \hbar \right)
\end{equation}%
satisfies (\ref{U(1)_symmetry}) if and only if the theory is self-dual.

Let us first note that the perturbative construction of $\mathcal{S}\left[
\phi ,\overline{\phi }\right] $ is encoded in the representation%
\begin{equation}
\mathcal{S}\left[ \phi ,\overline{\phi }\right] =\exp O\left[ \frac{\delta }{%
\delta \phi },\frac{\delta }{\delta \overline{\phi }}\right] \exp \left( 
\mathrm{i}S_{int}\left[ \phi ,\overline{\phi }\right] \right)
\label{normal_symbol_S}
\end{equation}%
where the differential operator in the functional derivatives%
\begin{eqnarray}
O\left[ \frac{\delta }{\delta \phi },\frac{\delta }{\delta \overline{\phi }}%
\right] &=&-\mathrm{i}\frac{\delta }{\delta \phi }\cdot \frac{\delta }{%
\delta \phi }-\mathrm{i}\frac{\delta }{\delta \overline{\phi }}\cdot \frac{%
\delta }{\delta \overline{\phi }}  \label{operator_O} \\
&&+\mathrm{i}\frac{\delta }{\delta \phi }\cdot \Delta \cdot \frac{\delta }{%
\delta \overline{\phi }}
\end{eqnarray}%
with%
\begin{equation}
\Delta _{AB\overset{.}{C}\overset{.}{D}}\left( x,y\right) =\int \frac{%
\mathrm{d}^{4}p}{\left( 2\pi \right) ^{4}}\mathrm{e}^{-\mathrm{i}p\cdot
\left( x-y\right) }\frac{\left[ p_{A\overset{.}{C}}p_{B\overset{.}{D}}+p_{A%
\overset{.}{D}}p_{B\overset{.}{C}}\right] }{p^{2}+\mathrm{i}0}
\end{equation}%
implements the Wick theorem with propagators (\ref{propagator_mixed}), (\ref%
{propagator_phi}) and (\ref{propagator_phi_bar}). Here and in what follows
we use condensed notation, e.g. 
\begin{eqnarray}
\Delta \cdot \frac{\delta }{\delta \overline{\phi }} &\equiv &\int \mathrm{d}%
^{4}y\Delta _{AB\overset{.}{C}\overset{.}{D}}\left( x,y\right) \frac{\delta 
}{\delta \overline{\phi }_{\overset{.}{C}\overset{.}{D}}\left( y\right) }, \\
\frac{\delta }{\delta \phi }\cdot \Delta &=&\int \mathrm{d}^{4}x\mathrm{d}%
^{4}y\frac{\delta }{\delta \phi _{AB}\left( y\right) }\Delta _{AB\overset{.}{%
C}\overset{.}{D}}\left( y,x\right) , \\
\frac{\delta }{\delta \phi }\cdot \Delta \cdot \frac{\delta }{\delta 
\overline{\phi }} &\equiv &\int \mathrm{d}^{4}x\mathrm{d}^{4}y\frac{\delta }{%
\delta \phi _{AB}\left( x\right) }\Delta _{AB\overset{.}{C}\overset{.}{D}%
}\left( x,y\right) \frac{\delta }{\delta \overline{\phi }_{\overset{.}{C}%
\overset{.}{D}}\left( y\right) }, \\
\frac{\delta }{\delta \phi }\cdot \frac{\delta }{\delta \phi } &\equiv &\int 
\mathrm{d}^{4}x\frac{\delta }{\delta \phi _{AB}\left( x\right) }\varepsilon
_{AC}\varepsilon _{BD}\frac{\delta }{\delta \phi _{CD}\left( x\right) },
\end{eqnarray}%
etc.\footnote{%
The helicity consrevation condition (\ref{U(1)_symmetry}) reads within this
notation%
\begin{equation*}
\phi \cdot \frac{\delta \mathcal{T}^{tree}}{\delta \phi }-\overline{\phi }%
\cdot \frac{\delta \mathcal{T}^{tree}}{\delta \overline{\phi }}=0.
\end{equation*}%
}. We also tacitly assume some implicit UV\ regularization\footnote{%
We also assume that $S_{int}$ contains all the necessary counterterms. Note
however, that these are of the order $O(\hbar )$ and higher and can be
effectively set to zero in what follows when we restrict ourselves to the
tree-level.}, typically the dimensional regularization\footnote{%
In this case, in order to preserve the four-dimensional spinor algebra, we
assume a dimensional reduction scheme.}. The details of this regularization
are not essential when we restrict ourselves to the tree-level. We get then
using (\ref{T_matrix})%
\begin{eqnarray}
\frac{\delta \mathcal{T}[\phi ,\overline{\phi }]}{\delta \phi _{AB}} &=&%
\mathrm{e}^{-\mathrm{i}\mathcal{T}[\phi ,\overline{\phi }]}\mathrm{e}^{O%
\left[ \frac{\delta }{\delta \phi },\frac{\delta }{\delta \overline{\phi }}%
\right] }\frac{\delta S_{int}[\phi ,\overline{\phi }]}{\delta \phi _{AB}}%
\mathrm{e}^{\mathrm{i}S_{int}[\phi ,\overline{\phi }]}  \notag \\
&=&\mathrm{e}^{-\mathrm{i}\mathcal{T}[\phi ,\overline{\phi }]}\mathrm{e}^{%
\left[ O_{\phi }+O_{\chi }+O_{\chi \phi }\right] }\frac{\delta S_{int}[\phi ,%
\overline{\phi }]}{\delta \phi _{AB}}\mathrm{e}^{\mathrm{i}S_{int}[\chi ,%
\overline{\chi }]}|_{\chi =\phi ,\overline{\chi }=\overline{\phi }}
\end{eqnarray}%
where we have doubled the number of fields in order to separate the action
of the operator $\mathrm{e}^{O\left[ \frac{\delta }{\delta \phi },\frac{%
\delta }{\delta \overline{\phi }}\right] }$ on both factors and introduced
commuting operators%
\begin{eqnarray}
O_{\phi } &=&O\left[ \frac{\delta }{\delta \phi },\frac{\delta }{\delta 
\overline{\phi }}\right] ,~~~O_{\chi }=O\left[ \frac{\delta }{\delta \chi },%
\frac{\delta }{\delta \overline{\chi }}\right] \\
O_{\chi \phi } &=&-2\mathrm{i}\frac{\delta }{\delta \phi }\cdot \frac{\delta 
}{\delta \chi }-2\mathrm{i}\frac{\delta }{\delta \overline{\phi }}\cdot 
\frac{\delta }{\delta \overline{\chi }}  \notag \\
&&+\mathrm{i}\frac{\delta }{\delta \phi }\cdot \Delta \cdot \frac{\delta }{%
\delta \overline{\chi }}+\mathrm{i}\frac{\delta }{\delta \chi }\cdot \Delta
\cdot \frac{\delta }{\delta \overline{\phi }}.
\end{eqnarray}%
Acting now with the diagonal operators $\mathrm{e}^{O_{\phi }}$ and $\mathrm{%
e}^{O_{\chi }}$on $\delta S_{int}[\phi ,\overline{\phi }]/\delta \phi _{AB}$
and $\mathrm{e}^{\mathrm{i}S_{int}[\chi ,\overline{\chi }]}$ respectively we
get%
\begin{equation}
\frac{\delta \mathcal{T}[\phi ,\overline{\phi }]}{\delta \phi _{AB}}=\mathrm{%
e}^{-\mathrm{i}\mathcal{T}[\phi ,\overline{\phi }]}\mathrm{e}^{O_{\chi \phi
}}\frac{\delta S_{int}[\phi ,\overline{\phi }]}{\delta \phi _{AB}}\mathrm{e}%
^{\mathrm{i}\mathcal{T}[\chi ,\overline{\chi }]}|_{\chi =\phi ,\overline{%
\chi }=\overline{\phi }}.
\end{equation}%
Note that within dimensional regularization%
\begin{equation}
\mathrm{e}^{O_{\phi }}\frac{\delta S_{int}[\phi ,\overline{\phi }]}{\delta
\phi _{AB}}=0  \label{tadpoles}
\end{equation}%
because $\delta S_{int}[\phi ,\overline{\phi }]/\delta \phi _{AB}$ is local
and therefore $O_{\phi }$ generates massless tadpoles. Equivalently we can
send the left hand side of (\ref{tadpoles}) effectively to zero when we are
interested in tree-level graphs only. The action of the operator $\mathrm{e}%
^{O_{\chi \phi }}$ on $\delta S_{int}[\phi ,\overline{\phi }]/\delta \phi
_{AB}$ shifts its functional arguments according to%
\begin{equation}
\phi \rightarrow \phi -2\mathrm{i}\frac{\delta }{\delta \chi }+\mathrm{i}%
\Delta \cdot \frac{\delta }{\delta \overline{\chi }},~~~~\overline{\phi }%
\rightarrow \overline{\phi }-2\mathrm{i}\frac{\delta }{\delta \overline{\chi 
}}+\mathrm{i}\frac{\delta }{\delta \chi }\cdot \Delta
\end{equation}%
and commuting the functional $\mathrm{e}^{\mathrm{i}\mathcal{T}[\chi ,%
\overline{\chi }]}$ (treated as an operator) with functional derivatives $%
\delta /\delta \chi $ and $\delta /\delta \overline{\chi }$ shifts these
derivatives%
\begin{equation}
\frac{\delta }{\delta \chi }\mathrm{e}^{\mathrm{i}\mathcal{T}[\chi ,%
\overline{\chi }]}=\mathrm{e}^{\mathrm{i}\mathcal{T}[\chi ,\overline{\chi }%
]}\left( \frac{\delta }{\delta \chi }+\mathrm{i}\frac{\delta \mathcal{T}%
[\chi ,\overline{\chi }]}{\delta \chi }\right) .
\end{equation}%
As a result of this operations we get $\delta \mathcal{T}/\delta \phi _{AB}$
in the form of an action of the differential operation on trivial functional 
$F[\chi ,\overline{\chi }]=1$ 
\begin{equation}
\frac{\delta \mathcal{T}}{\delta \phi _{AB}}=\frac{\delta S_{int}}{\delta
\phi _{AB}}\left[ \mathcal{J},\overline{\mathcal{J}}\right] 1|_{\chi =\phi ,%
\overline{\chi }=\overline{\phi }}  \label{dT/dphi_complet}
\end{equation}%
and similarly%
\begin{equation}
\frac{\delta \mathcal{T}}{\delta \overline{\phi }_{\overset{.}{A}\overset{.}{%
B}}}=\frac{\delta S_{int}}{\delta \overline{\phi }_{\overset{.}{A}\overset{.}%
{B}}}\left[ \mathcal{J},\overline{\mathcal{J}}\right] 1|_{\chi =\phi ,%
\overline{\chi }=\overline{\phi }}  \label{dT/dphibar_complet}
\end{equation}%
where $\mathcal{J}$ and $\overline{\mathcal{J}}$ are differential operators
in functional derivatives given by%
\begin{eqnarray}
\mathcal{J} &=&\phi +2\frac{\delta \mathcal{T}[\chi ,\overline{\chi }]}{%
\delta \chi }-\Delta \cdot \frac{\delta \mathcal{T}[\chi ,\overline{\chi }]}{%
\delta \overline{\chi }}+\hbar \frac{\delta }{\delta \chi }+\mathrm{i}\hbar
\Delta \cdot \frac{\delta }{\delta \overline{\chi }} \\
\overline{\mathcal{J}} &=&\overline{\phi }+2\frac{\delta \mathcal{T}[\chi ,%
\overline{\chi }]}{\delta \overline{\chi }}-\frac{\delta \mathcal{T}[\chi ,%
\overline{\chi }]}{\delta \chi }\cdot \Delta +\hbar \frac{\delta }{\delta 
\overline{\chi }}+\mathrm{i}\hbar \frac{\delta }{\delta \chi }\cdot \Delta .
\end{eqnarray}%
In these formulas we restored the dependence on the Planck constant. At tree
level therefore, writing%
\begin{equation}
S_{int}\left[ \phi ,\overline{\phi }\right] =\int \mathrm{d}^{4}x\mathcal{L}%
_{int}\left( \phi \left( x\right) ^{2},\overline{\phi }\left( x\right)
^{2}\right) ,
\end{equation}%
and taking the leading terms in the expansion in $\hbar $ on both sides of (%
\ref{dT/dphi_complet}), (\ref{dT/dphibar_complet}) we get 
\begin{eqnarray}
\frac{\delta \mathcal{T}^{tree}}{\delta \phi _{AB}} &=&\frac{\delta S_{int}}{%
\delta J_{AB}}[J,\overline{J}]=2J^{AB}\frac{\partial \mathcal{L}_{int}(J^{2},%
\overline{J}^{2})}{\partial J^{2}},  \label{dT/dphi} \\
\frac{\delta \mathcal{T}^{tree}}{\delta \overline{\phi }_{\overset{.}{A}%
\overset{.}{B}}} &=&\frac{\delta S_{int}}{\delta \overline{J}_{\overset{.}{A}%
\overset{.}{B}}}[J,\overline{J}]=2\overline{J}^{\overset{.}{A}\overset{.}{B}}%
\frac{\partial \mathcal{L}_{int}(J^{2},\overline{J}^{2})}{\partial \overline{%
J}^{2}}.  \label{dT/dphibar}
\end{eqnarray}%
Here we denoted (using our condensed notation)%
\begin{eqnarray}
J &=&\mathcal{J}|_{\hbar \rightarrow 0,~\chi =\phi ,\overline{\chi }=%
\overline{\phi }}=\phi +2\frac{\delta \mathcal{T}^{tree}}{\delta \phi }%
-\Delta \cdot \frac{\delta \mathcal{T}^{tree}}{\delta \overline{\phi }}
\label{J} \\
\overline{J} &=&\overline{\mathcal{J}}|_{\hbar \rightarrow 0,~\chi =\phi ,%
\overline{\chi }=\overline{\phi }}=\overline{\phi }+2\frac{\delta \mathcal{T}%
^{tree}}{\delta \overline{\phi }}-\frac{\delta \mathcal{T}^{tree}}{\delta
\phi }\cdot \Delta  \label{Jbar}
\end{eqnarray}%
Note that the right hand sides of (\ref{dT/dphi}) and (\ref{dT/dphibar})\
are local when expressed in terms of the variables $J$ and $\overline{J}$.
From (\ref{dT/dphi}) and (\ref{dT/dphibar}) it follows 
\begin{equation}
\phi \cdot \frac{\delta \mathcal{T}^{tree}}{\delta \phi }-\overline{\phi }%
\cdot \frac{\delta \mathcal{T}^{tree}}{\delta \overline{\phi }}=\int \mathrm{%
d}^{4}x\left[ 2\phi _{AB}J^{AB}\frac{\partial \mathcal{L}_{int}}{\partial
J^{2}}-2\overline{\phi }_{\overset{.}{A}\overset{.}{B}}\overline{J}^{\overset%
{.}{A}\overset{.}{B}}\frac{\partial \mathcal{L}_{int}}{\partial \overline{J}%
^{2}}\right] .  \label{T_transformation}
\end{equation}%
Using (\ref{J}) and (\ref{Jbar}) and with help of \ (\ref{dT/dphi}) and (\ref%
{dT/dphibar}) we can express $\phi _{AB}$ as a functional of $J_{AB}$ and $%
\overline{J}_{\overset{.}{A}\overset{.}{B}}$%
\begin{eqnarray}
\phi _{AB} &=&J-2\frac{\delta \mathcal{T}^{tree}}{\delta \phi }+\Delta \cdot 
\frac{\delta \mathcal{T}^{tree}}{\delta \overline{\phi }}  \notag \\
&=&J-4J\frac{\partial \mathcal{L}_{int}}{\partial J^{2}}+2\Delta \cdot 
\overline{J}\frac{\partial \mathcal{L}_{int}}{\partial \overline{J}^{2}}
\end{eqnarray}%
and similarly%
\begin{equation}
\overline{\phi }=\overline{J}-4J\frac{\partial \mathcal{L}_{int}}{\partial 
\overline{J}^{2}}+2\frac{\partial \mathcal{L}_{int}}{\partial J^{2}}J\cdot
\Delta
\end{equation}%
and with help of (\ref{T_transformation})%
\begin{eqnarray}
&&\phi \cdot \frac{\delta \mathcal{T}^{tree}}{\delta \phi }-\overline{\phi }%
\cdot \frac{\delta \mathcal{T}^{tree}}{\delta \overline{\phi }}  \notag \\
&=&\int \mathrm{d}^{4}x\left[ 2J^{2}\frac{\partial \mathcal{L}_{int}}{%
\partial J^{2}}-2\overline{J}^{2}\frac{\partial \mathcal{L}_{int}}{\partial 
\overline{J}^{2}}-8J^{2}\left( \frac{\partial \mathcal{L}_{int}}{\partial
J^{2}}\right) ^{2}+8\overline{J}^{2}\left( \frac{\partial \mathcal{L}_{int}}{%
\partial \overline{J}^{2}}\right) ^{2}\right]  \label{T_transform}
\end{eqnarray}%
Note that the right hand side is local again when expressed in terms of $J$
and $\overline{J}$: the nonlocal terms containing the mixed propagator $%
\Delta _{\overset{.}{A}\overset{.}{B}AB}$\ completely canceled each other.
Inserting now (cf (\ref{L_in_phi_phibar})) 
\begin{equation}
\frac{\partial \mathcal{L}_{int}}{\partial J^{2}}=\frac{1}{8}+\frac{\partial 
\mathcal{L}}{\partial J^{2}},~~~~~\frac{\partial \mathcal{L}_{int}}{\partial 
\overline{J}^{2}}=\frac{1}{8}+\frac{\partial \mathcal{L}}{\partial \overline{%
J}^{2}}
\end{equation}%
into (\ref{T_transform}) we get finally 
\begin{equation}
\phi \cdot \frac{\delta \mathcal{T}^{tree}}{\delta \phi }-\overline{\phi }%
\cdot \frac{\delta \mathcal{T}^{tree}}{\delta \overline{\phi }}=\int \mathrm{%
d}^{4}x\left[ -8J^{2}\left( \frac{\partial \mathcal{L}}{\partial J^{2}}%
\right) ^{2}+8\overline{J}^{2}\left( \frac{\partial \mathcal{L}}{\partial 
\overline{J}^{2}}\right) ^{2}+\frac{1}{8}J^{2}-\frac{1}{8}\overline{J}^{2}%
\right]  \label{duality_iff_helicity_conservation}
\end{equation}%
But the on the right hand side of this equation we recognize the NGZ
constraint (\ref{NGZ}) vanishing of which is the necessary and sufficient
condition for self-dual theories. Therefore at the tree-level the helicity
is conserved if and only if the theory is self-dual.

\section{Normal ordering and modified Feynman rules}

Writing in the formula (\ref{normal_symbol_S}) for the normal symbol of the
S-matrix\footnote{%
Note that the individual terms in (\ref{operator_O}) are all commuting
operators.} 
\begin{equation}
\exp O\left[ \frac{\delta }{\delta \phi },\frac{\delta }{\delta \overline{%
\phi }}\right] =\exp \left( \mathrm{i}\frac{\delta }{\delta \phi }\cdot
\Delta \cdot \frac{\delta }{\delta \overline{\phi }}\right) \exp \left( -%
\mathrm{i}\frac{\delta }{\delta \phi }\cdot \frac{\delta }{\delta \phi }-%
\mathrm{i}\frac{\delta }{\delta \overline{\phi }}\cdot \frac{\delta }{\delta 
\overline{\phi }}\right) ,
\end{equation}%
we can rearrange the calculation of $\mathcal{S}\left[ \phi ,\overline{\phi }%
\right] $ as 
\begin{equation}
\mathcal{S}\left[ \phi ,\overline{\phi }\right] =\exp \left( \mathrm{i}\frac{%
\delta }{\delta \phi }\cdot \Delta \cdot \frac{\delta }{\delta \overline{%
\phi }}\right) \exp \left( \mathrm{i}S_{int}^{NO}\left[ \phi ,\overline{\phi 
}\right] \right) .  \label{normal_symbol_S_NO}
\end{equation}%
Here the normal ordered\footnote{%
The term \textquotedblleft normal ordered" here has not to be confused with
the usual operator normal ordering with respect to the creation and
annihilation operators.} interaction action $S_{int}^{NO}\left[ \phi ,%
\overline{\phi }\right] $ is defined as%
\begin{equation}
\exp \left( \mathrm{i}S_{int}^{NO}\left[ \phi ,\overline{\phi }\right]
\right) =\exp \left( -\mathrm{i}\frac{\delta }{\delta \phi }\cdot \frac{%
\delta }{\delta \phi }-\mathrm{i}\frac{\delta }{\delta \overline{\phi }}%
\cdot \frac{\delta }{\delta \overline{\phi }}\right) \exp \left( \mathrm{i}%
S_{int}\left[ \phi ,\overline{\phi }\right] \right) .  \label{S_int_NO}
\end{equation}%
Because the operator 
\begin{equation}
O_{local}\equiv -\mathrm{i}\frac{\delta }{\delta \phi }\cdot \frac{\delta }{%
\delta \phi }-\mathrm{i}\frac{\delta }{\delta \overline{\phi }}\cdot \frac{%
\delta }{\delta \overline{\phi }}
\end{equation}%
is local and does not generate space-time derivatives of $\phi ,\overline{%
\phi }$, the functional $S_{int}^{NO}\left[ \phi ,\overline{\phi }\right] $
is also local, i.e.%
\begin{equation}
S_{int}^{NO}\left[ \phi ,\overline{\phi }\right] =\int \mathrm{d}^{4}x%
\mathcal{L}_{int}^{NO}\left[ \phi ,\overline{\phi }\right] ,
\end{equation}%
and $\mathcal{L}_{int}\left[ \phi ,\overline{\phi }\right] $ is a function
of the invariants $\phi ^{2}$ and $\overline{\phi }^{2}$only. Provided the
normal ordered interaction action $S_{int}^{NO}\left[ \phi ,\overline{\phi }%
\right] $ is known, we can calculate the $S-$matrix equivalently using
formula (\ref{normal_symbol_S_NO}), i.e. using only the mixed propagator $%
\langle T\phi \overline{\phi }\rangle $ for the internal lines. This
approach is much more effective and also more physical because only the
mixed propagator has the one-particle pole. The contributions of the contact
propagator terms are naturally accumulated in the normal ordered interaction
vertices derived from $\mathcal{L}_{int}^{NO}\left[ \phi ,\overline{\phi }%
\right] $.

Note that the equation (\ref{S_int_NO}) has the same structure as formula (%
\ref{normal_symbol_S}). Therefore, $S_{int}^{NO}\left[ \phi ,\overline{\phi }%
\right] $ is represented with connected graphs generated by the Wick
contractions encoded in the operator $\exp O_{local}$. Moreover, because the
operator $O_{local}$ generates local contractions, the loop graphs are
proportional to $\delta ^{\left( 4\right) }\left( 0\right) $, which vanish
in dimensional regularization. So that within dimensional regularization,
which we will implicitly assume in what follows, only the tree graphs
contribute. We can therefore start with (\ref{S_int_NO}) and repeat all the
manipulations which lead us from (\ref{normal_symbol_S}) to equations (\ref%
{dT/dphi}) and (\ref{dT/dphibar}). The results is%
\begin{eqnarray}
\frac{\delta S_{int}^{NO}\left[ \phi ,\overline{\phi }\right] }{\delta \phi
_{AB}} &=&\frac{\partial \mathcal{L}_{int}^{NO}\left[ \phi ,\overline{\phi }%
\right] }{\partial \phi _{AB}}=2J^{AB}\frac{\partial \mathcal{L}_{int}(J^{2},%
\overline{J}^{2})}{\partial J^{2}}, \\
\frac{\delta S_{int}^{NO}\left[ \phi ,\overline{\phi }\right] }{\delta 
\overline{\phi }_{\overset{.}{A}\overset{.}{B}}} &=&\frac{\partial \mathcal{L%
}_{int}^{NO}\left[ \phi ,\overline{\phi }\right] }{\partial \overline{\phi }%
_{\overset{.}{A}\overset{.}{B}}}=2\overline{J}^{\overset{.}{A}\overset{.}{B}}%
\frac{\partial \mathcal{L}_{int}(J^{2},\overline{J}^{2})}{\partial \overline{%
J}^{2}},
\end{eqnarray}%
in this case with%
\begin{equation}
J=\phi +2\frac{\partial \mathcal{L}_{int}^{NO}\left[ \phi ,\overline{\phi }%
\right] }{\partial \phi },~~~~~\overline{J}=\overline{\phi }+2\frac{\partial 
\mathcal{L}_{int}^{NO}\left[ \phi ,\overline{\phi }\right] }{\partial 
\overline{\phi }}.
\end{equation}%
Expressing now $\phi $ and $\overline{\phi }$ in terms of $J$ and $\overline{%
J}$ we get therefore an analogue of (\ref{T_transform}) and finally an
analogue of (\ref{duality_iff_helicity_conservation}), again with the NGZ
constraint on the right hand side: 
\begin{equation*}
\phi _{AB}\frac{\partial \mathcal{L}_{int}^{NO}\left[ \phi ,\overline{\phi }%
\right] }{\partial \phi _{AB}}-\overline{\phi }_{\overset{.}{A}\overset{.}{B}%
}\frac{\partial \mathcal{L}_{int}^{NO}\left[ \phi ,\overline{\phi }\right] }{%
\partial \overline{\phi }_{\overset{.}{A}\overset{.}{B}}}=-8J^{2}\left( 
\frac{\partial \mathcal{L}}{\partial J^{2}}\right) ^{2}+8\overline{J}%
^{2}\left( \frac{\partial \mathcal{L}}{\partial \overline{J}^{2}}\right)
^{2}+\frac{1}{8}J^{2}-\frac{1}{8}\overline{J}^{2}.
\end{equation*}%
We can thus conclude that the theory is self-dual if and only if the
corresponding normal ordered interaction Lagrangian $\mathcal{L}_{int}^{NO}%
\left[ \phi ,\overline{\phi }\right] $ is $U(1)$ invariant, i.e. provided
its dependence on $\phi $ and $\overline{\phi }$ is through the combination $%
\phi ^{2}\overline{\phi }^{2}$ only: $\mathcal{L}_{int}^{NO}\left[ \phi ,%
\overline{\phi }\right] =L\left( \phi ^{2}\overline{\phi }^{2}\right) $.

\subsection{Helicity conservation at higher loops}

The latter statement allows us to enlarge the validity of the conclusion of
the previous section concerning tree-level helicity conservation to all
higher loop graphs with vertices derived from self-dual Lagrangian of the type (\ref{L_in_phi_phibar})
satisfying the NGZ condition. Indeed, note that in the formula (\ref%
{normal_symbol_S_NO}) the operator $\exp \left( \mathrm{i}\frac{\delta }{%
\delta \phi }\cdot \Delta \cdot \frac{\delta }{\delta \overline{\phi }}%
\right) $ preserves helicity. Therefore provided the theory is self-dual,
the corresponding normal ordered interaction Lagrangian $\mathcal{L}%
_{int}^{NO}\left[ \phi ,\overline{\phi }\right] $ is $U(1)$ invariant, and
therefore so must be the $S-$matrix $\mathcal{S}\left[ \phi ,\overline{\phi }%
\right] $. This implies the helicity conservation. Of course, here we assume
that the implicit regularization does not violate the $U\left( 1\right) $
symmetry and therefore only $U\left( 1\right) $ symmetric counterterms are
needed.

Remarkably, this can be formally understood also on the Lagrangian level.
Note that the mixed propagator $\langle T\phi \overline{\phi }\rangle =%
\mathrm{i}\Delta $ satisfies%
\begin{eqnarray}
&&\int \mathrm{d}^{4}z~\Delta _{AB\overset{.}{C}\overset{.}{D}}\left(
x-z\right) \varepsilon ^{\overset{.}{C}\overset{.}{G}}\varepsilon ^{\overset{%
.}{D}\overset{.}{H}}\Delta _{EF\overset{.}{G}\overset{.}{H}}\left( z-y\right)
\notag \\
&=&2\delta ^{\left( 4\right) }\left( x-y\right) \left[ \varepsilon
_{AE}\varepsilon _{BF}+\varepsilon _{AF}\varepsilon _{BE}\right] \\
&&\int \mathrm{d}^{4}z~\Delta _{AB\overset{.}{C}\overset{.}{D}}\left(
x-z\right) \varepsilon ^{AE}\varepsilon ^{BF}\Delta _{EF\overset{.}{G}%
\overset{.}{H}}\left( z-y\right) .  \notag \\
&=&2\delta ^{\left( 4\right) }\left( x-y\right) \left[ \varepsilon _{\overset%
{.}{C}\overset{.}{G}}\varepsilon _{\overset{.}{D}\overset{.}{H}}+\varepsilon
_{\overset{.}{C}\overset{.}{H}}\varepsilon _{\overset{.}{C}\overset{.}{G}}%
\right]
\end{eqnarray}%
Therefore, introducing formal functional Gaussian integration 
\begin{equation}
\exp \left( \mathrm{i}\frac{\delta }{\delta \phi }\cdot \Delta \cdot \frac{%
\delta }{\delta \overline{\phi }}\right) =\int D\varphi D\overline{\varphi }%
\exp \left( \frac{\mathrm{i}}{4}\varphi \cdot \Delta \cdot \overline{\varphi 
}+\varphi \cdot \frac{\delta }{\delta \phi }+\overline{\varphi }\cdot \frac{%
\delta }{\delta \overline{\phi }}\right) ,
\end{equation}%
we can represent the right hand side of the formula (\ref{normal_symbol_S_NO}%
) as%
\begin{eqnarray}
\mathcal{S}\left[ \phi ,\overline{\phi }\right] &=&\int D\varphi D\overline{%
\varphi }\exp \left( \frac{\mathrm{i}}{4}\varphi \cdot \Delta \cdot 
\overline{\varphi }+\varphi \cdot \frac{\delta }{\delta \phi }+\overline{%
\varphi }\cdot \frac{\delta }{\delta \overline{\phi }}\right) \exp \left( 
\mathrm{i}S_{int}^{NO}\left[ \phi ,\overline{\phi }\right] \right)  \notag \\
&=&\int D\varphi D\overline{\varphi }\exp \left( \frac{\mathrm{i}}{4}\varphi
\cdot \Delta \cdot \overline{\varphi }\right) \exp \left( \mathrm{i}%
S_{int}^{NO}\left[ \phi +\varphi ,\overline{\phi }+\overline{\varphi }\right]
\right)  \notag \\
&=&\int D\varphi D\overline{\varphi }\exp \left( \frac{\mathrm{i}}{4}\left(
\varphi -\phi \right) \cdot \Delta \cdot \left( \overline{\varphi }-%
\overline{\phi }\right) \right) \exp \left( \mathrm{i}S_{int}^{NO}\left[
\varphi ,\overline{\varphi }\right] \right) .
\end{eqnarray}%
The latter formula formally corresponds to the functional integral
representation of the $S-$matrix in theory with \textquotedblleft classical
normal ordered action\textquotedblleft\ $S^{NO}\left[ \varphi ,\overline{%
\varphi }\right] $ of the form 
\begin{equation}
S^{NO}\left[ \varphi ,\overline{\varphi }\right] =\frac{1}{4}\varphi \cdot
\Delta \cdot \overline{\varphi }+S_{int}^{NO}\left[ \varphi ,\overline{%
\varphi }\right] ,
\end{equation}%
which is formulated solely in terms of gauge invariant fileds $\varphi $ and 
$\overline{\varphi }$ without necessity to relate it to the potential $%
A_{\mu }$. This is in contrast to the original action $S\left[ \varphi ,%
\overline{\varphi }\right] $%
\begin{equation}
S\left[ \varphi ,\overline{\varphi }\right] =-\frac{1}{8}\varphi ^{2}-\frac{1%
}{8}\overline{\varphi }^{2}+S_{int}\left[ \varphi ,\overline{\varphi }\right]
.
\end{equation}%
for which the path integral quantization needs $\varphi $ and $\overline{%
\varphi }$ to be expressed in terms of $A_{\mu }$ and a gauge fixing term
has to be added.

For self-dual theories the action $S^{NO}\left[ \varphi ,\overline{\varphi }\right] $\ shows
manifest $U\left( 1\right) $ symmetry and implies therefore also manifestly
the helicity conservation. Note however that the kinetic term of this action
is non-local, nevertheless it generates formally the right propagator $%
\langle T\phi \overline{\phi }\rangle =\mathrm{i}\Delta $. This is the price
we pay for working directly with the variables $\varphi $ and $\overline{%
\varphi }$.

\subsection{Calculation of the normal ordered Lagrangian}

According to the definition (\ref{S_int_NO}), the normal ordered interaction
Lagrangian $\mathcal{L}_{int}^{NO}\left[ \phi ,\overline{\phi }\right] $ can
be obtained as a sum of connected graphs with vertices from $S_{int}\left[
\phi ,\overline{\phi }\right] $ and internal lines corresponding only to the
contact propagators $\langle T\phi \phi \rangle $ and $\langle T\overline{%
\phi }\overline{\phi }\rangle $ (see (\ref{propagator_phi}) and (\ref%
{propagator_phi_bar})). Due to the locality, the loops are proportional to $%
\delta ^{\left( 4\right) }\left( 0\right) $ which vanishes in the
dimensional regularization and thus only the tree graphs are relevant. These
can be summed up as follows. Let us rewrite (\ref{S_int_NO}) in the form%
\begin{eqnarray}
\exp \left( \mathrm{i}S_{int}^{NO}\left[ \phi ,\overline{\phi }\right]
\right) &=&\int D\varphi D\overline{\varphi }\mathrm{e}^{\left( -\frac{%
\mathrm{i}}{4}\varphi \cdot \varphi -\frac{\mathrm{i}}{4}\overline{\varphi }%
\cdot \overline{\varphi }+\varphi \cdot \frac{\delta }{\delta \phi }+%
\overline{\varphi }\cdot \frac{\delta }{\delta \overline{\phi }}\right) }%
\mathrm{e}^{\mathrm{i}S_{int}\left[ \phi ,\overline{\phi }\right] }  \notag
\\
&=&\int D\varphi D\overline{\varphi }\mathrm{e}^{\left( -\frac{\mathrm{i}}{4}%
\varphi \cdot \varphi -\frac{\mathrm{i}}{4}\overline{\varphi }\cdot 
\overline{\varphi }\right) }\mathrm{e}^{\mathrm{i}S_{int}\left[ \phi
+\varphi ,\overline{\phi }+\overline{\varphi }\right] }  \notag \\
&=&\int D\varphi D\overline{\varphi }\mathrm{e}^{\left( -\frac{\mathrm{i}}{4}%
\left( \varphi -\phi \right) \cdot \left( \varphi -\phi \right) -\frac{%
\mathrm{i}}{4}\left( \overline{\varphi }-\overline{\phi }\right) \cdot
\left( \overline{\varphi }-\overline{\phi }\right) +\mathrm{i}S_{int}\left[
\varphi ,\overline{\varphi }\right] \right) }.  \notag \\
&&
\end{eqnarray}%
The result of the tree-level calculation of the functional integral then
corresponds to%
\begin{equation}
S_{int}^{NO}\left[ \phi ,\overline{\phi }\right] =S_{int}\left[ \varphi ,%
\overline{\varphi }\right] -\frac{1}{4}\left( \varphi -\phi \right) \cdot
\left( \varphi -\phi \right) -\frac{1}{4}\left( \overline{\varphi }-%
\overline{\phi }\right) \cdot \left( \overline{\varphi }-\overline{\phi }%
\right) ,  \label{generating_lagrangian}
\end{equation}%
where $\varphi ,\overline{\varphi }$ satisfy the classical equation of
motion 
\begin{equation}
-\frac{1}{2}\left( \varphi -\phi \right) _{AB}+\frac{\partial \mathcal{L}%
_{int}}{\partial \varphi ^{AB}}=0,~~~~-\frac{1}{2}\left( \overline{\varphi }-%
\overline{\phi }\right) _{\overset{\cdot }{A}\overset{\cdot }{B}}+\frac{%
\partial \mathcal{L}_{int}}{\partial \overline{\varphi }^{\overset{\cdot }{A}%
\overset{\cdot }{B}}}=0 .  \label{algebraic_EOM}
\end{equation}%
Therefore%
\begin{equation}
\mathcal{L}_{int}^{NO}\left[ \phi ,\overline{\phi }\right] =\mathcal{L}_{int}%
\left[ \varphi ,\overline{\varphi }\right] -\frac{\partial \mathcal{L}_{int}%
}{\partial \varphi ^{AB}}\frac{\partial \mathcal{L}_{int}}{\partial \varphi
_{AB}}-\frac{\partial \mathcal{L}_{int}}{\partial \overline{\varphi }^{%
\overset{\cdot }{A}\overset{\cdot }{B}}}\frac{\partial \mathcal{L}_{int}}{%
\partial \overline{\varphi }_{\overset{\cdot }{A}\overset{\cdot }{B}}}.
\label{LNO_mod}
\end{equation}%
and $\varphi ,\overline{\varphi }$ are solutions of (\ref{algebraic_EOM}).
Note that both $\mathcal{L}_{int}\left[ \varphi ,\overline{\varphi }\right] $
and $\mathcal{L}_{int}^{NO}\left[ \varphi ,\overline{\varphi }\right] $ are
functions of the invariants $\varphi ^{2}$ and $\overline{\varphi }^{2}$,
therefore%
\begin{equation}
\frac{\partial \mathcal{L}_{int}}{\partial \varphi ^{AB}}=2\varphi _{AB}%
\frac{\partial \mathcal{L}_{int}}{\partial \varphi ^{2}}.
\end{equation}%
Finally we rewrite (\ref{LNO_mod}) as%
\begin{equation}
\mathcal{L}_{int}^{NO}\left( \phi ^{2},\overline{\phi }^{2}\right) =\mathcal{%
L}_{int}\left( \varphi ^{2},\overline{\varphi }^{2}\right) -4\varphi
^{2}\left( \frac{\partial \mathcal{L}_{int}\left( \varphi ^{2},\overline{%
\varphi }^{2}\right) }{\partial \varphi ^{2}}\right) ^{2}-4\overline{\varphi 
}^{2}\left( \frac{\partial \mathcal{L}_{int}\left( \varphi ^{2},\overline{%
\varphi }^{2}\right) }{\partial \overline{\varphi }^{2}}\right) ^{2},
\label{L_int_general_final}
\end{equation}%
and the equations (\ref{algebraic_EOM}) can be written in the form 
\begin{equation}
\phi ^{2}=\varphi ^{2}\left( 1-4\frac{\partial \mathcal{L}_{int}\left(
\varphi ^{2},\overline{\varphi }^{2}\right) }{\partial \varphi ^{2}}\right)
^{2},~~~~\overline{\phi }^{2}=\overline{\varphi }^{2}\left( 1-4\frac{%
\partial \mathcal{L}_{int}\left( \varphi ^{2},\overline{\varphi }^{2}\right) 
}{\partial \overline{\varphi }^{2}}\right) ^{2}.  \label{EOM_general_final}
\end{equation}%
The summation of the tree graphs with the contact propagators $\langle T\phi
\phi \rangle $ and $\langle T\overline{\phi }\overline{\phi }\rangle $ is
therefore equivalent to the solution of the algebraic equations (\ref%
{EOM_general_final}) with respect to $\varphi ^{2}$ and $\overline{\varphi }%
^{2}$ and inserting then the solution into (\ref{L_int_general_final}).

Let us note, that the relation (\ref{S_int_NO}) connecting the original
Lagrangian with the normal ordered one is invertible, namely 
\begin{equation}
\exp \left( \mathrm{i}S_{int}\left[ \phi ,\overline{\phi }\right] \right)
=\exp \left( \mathrm{i}\frac{\delta }{\delta \phi }\cdot \frac{\delta }{%
\delta \phi }+\mathrm{i}\frac{\delta }{\delta \overline{\phi }}\cdot \frac{%
\delta }{\delta \overline{\phi }}\right) \exp \left( \mathrm{i}S_{int}^{NO}%
\left[ \phi ,\overline{\phi }\right] \right) .  \label{NO_inversion}
\end{equation}%
Repeating the above formal manipulation we can write the result of the
inversion as%
\begin{equation}
\mathcal{L}_{int}\left( \phi ^{2},\overline{\phi }^{2}\right) =\mathcal{L}%
_{int}^{NO}\left( \varphi ^{2},\overline{\varphi }^{2}\right) +4\varphi
^{2}\left( \frac{\partial \mathcal{L}_{int}^{NO}\left( \varphi ^{2},%
\overline{\varphi }^{2}\right) }{\partial \varphi ^{2}}\right) ^{2}+4%
\overline{\varphi }^{2}\left( \frac{\partial \mathcal{L}_{int}^{NO}\left(
\varphi ^{2},\overline{\varphi }^{2}\right) }{\partial \overline{\varphi }%
^{2}}\right) ^{2},  \label{L_int_inversion}
\end{equation}%
where now $\varphi ^{2}$ and $\overline{\varphi }^{2}$ are solution of
algebraic equations 
\begin{equation}
\phi ^{2}=\varphi ^{2}\left( 1+4\frac{\partial \mathcal{L}_{int}^{NO}\left(
\varphi ^{2},\overline{\varphi }^{2}\right) }{\partial \varphi ^{2}}\right)
^{2},~~~~\overline{\phi }^{2}=\overline{\varphi }^{2}\left( 1+4\frac{%
\partial \mathcal{L}_{int}^{NO}\left( \varphi ^{2},\overline{\varphi }%
^{2}\right) }{\partial \overline{\varphi }^{2}}\right) ^{2}.
\label{L_int_inversion_equation}
\end{equation}

Let us note, that the starting point for derivation of the equations (\ref%
{L_int_inversion})\_ and (\ref{L_int_inversion_equation}) is the analogue of
(\ref{generating_lagrangian}) and (\ref{algebraic_EOM}), namely%
\begin{equation}
S_{int}\left[ \phi ,\overline{\phi }\right] =S_{int}^{NO}\left[ \varphi ,%
\overline{\varphi }\right] +\frac{1}{4}\left( \varphi -\phi \right) \cdot
\left( \varphi -\phi \right) +\frac{1}{4}\left( \overline{\varphi }-%
\overline{\phi }\right) \cdot \left( \overline{\varphi }-\overline{\phi }%
\right) ,
\end{equation}%
where $\varphi ,\overline{\varphi }$ satisfy the classical equation of
motion 
\begin{equation}
\frac{1}{2}\left( \varphi -\phi \right) _{AB}+\frac{\partial \mathcal{L}%
_{int}^{NO}}{\partial \varphi ^{AB}}=0,~~~~\frac{1}{2}\left( \overline{%
\varphi }-\overline{\phi }\right) _{\overset{\cdot }{A}\overset{\cdot }{B}}+%
\frac{\partial \mathcal{L}_{int}^{NO}}{\partial \overline{\varphi }^{\overset%
{\cdot }{A}\overset{\cdot }{B}}}=0.
\end{equation}%
This can be directly compared with the auxiliary field construction of the
self-dual actions of Ivanov and Zupnik \cite{Ivanov:2003uj}, \cite%
{Ivanov:2012bq}.  Up to a different normalization of the fields, the normal
ordered action can be identified with the $U\left( 1\right) $ invariant
off-shell action from their construction where $\varphi ,\overline{\varphi }$
play the role of the auxiliary fields.

In the next two subsection we will illustrate the application of the
correspondence$\mathcal{L}_{int}\leftrightarrow \mathcal{L}_{int}^{NO}$ in
two special cases for which we can obtain the solution of both problems in a
closed form.

\subsection{Normal ordered form of the Born-Infeld Lagrangian}

As the first illustration, let us find the normal ordered form of the BI
Lagrangian. In the case of BI theory it is convenient to use the following
change of variables (first introduced in \cite{Bellucci:2000bd})%
\begin{equation}
\varphi _{AB}=\frac{1}{\sqrt{2}}\psi _{AB}\frac{1+\overline{\eta }}{1-\eta 
\overline{\eta }},~~~~~\overline{\varphi }_{\overset{\cdot }{A}\overset{%
\cdot }{B}}=\frac{1}{\sqrt{2}}\overline{\psi }_{\overset{\cdot }{A}\overset{%
\cdot }{B}}\frac{1+\eta }{1-\eta \overline{\eta }}
\end{equation}%
where%
\begin{equation}
\eta =\frac{\psi ^{2}}{16\Lambda ^{4}},~~~~~\overline{\eta }=\frac{\overline{%
\psi }^{2}}{16\Lambda ^{4}},
\end{equation}%
and therefore%
\begin{equation}
\varphi ^{2}=8\Lambda ^{4}\eta \left( \frac{1+\overline{\eta }}{1-\eta 
\overline{\eta }}\right) ^{2},~~~~~\overline{\varphi }^{2}=8\Lambda ^{4}%
\overline{\eta }\left( \frac{1+\eta }{1-\eta \overline{\eta }}\right) ^{2}
\end{equation}%
In \cite{Ferrara:2016crd} it was found that, when expressed in terms of
these new fields, the BI Lagrangian simplifies to rational function of $\eta 
$ and $\overline{\eta }$ 
\begin{equation}
\mathcal{L}_{BI}=-\Lambda ^{4}\frac{\eta +\overline{\eta }+2\eta \overline{%
\eta }}{1-\eta \overline{\eta }},
\end{equation}%
and the interaction part looks like%
\begin{equation}
\mathcal{L}_{BI,int}\equiv \mathcal{L}_{BI}+\frac{1}{8}\left( \varphi ^{2}+%
\overline{\varphi }^{2}\right) =-\Lambda ^{4}\eta \overline{\eta }\frac{%
\left( 1+\eta \right) \left( 1+\overline{\eta }\right) }{\left( 1-\eta 
\overline{\eta }\right) ^{2}}.
\end{equation}%
Inserting now the new parametrization into (\ref{L_int_general_final}) we get%
\begin{equation}
\mathcal{L}_{BI,int}^{NO}\left( \phi ^{2},\overline{\phi }^{2}\right)
=2\Lambda ^{4}\eta \overline{\eta }\frac{1+\eta \overline{\eta }}{\left(
1-\eta \overline{\eta }\right) ^{2}}
\end{equation}%
and the algebraic equation (\ref{EOM_general_final}) are transformed to 
\begin{equation}
\phi ^{2}=8\Lambda ^{4}\frac{\eta }{\left( 1-\eta \overline{\eta }\right)
^{2}},~\ ~~~\overline{\phi }^{2}=8\Lambda ^{4}\frac{\overline{\eta }}{\left(
1-\eta \overline{\eta }\right) ^{2}}
\end{equation}%
From the latter equation we get%
\begin{equation}
\phi ^{2}\overline{\phi }^{2}=64\Lambda ^{8}\frac{\eta \overline{\eta }}{%
\left( 1-\eta \overline{\eta }\right) ^{4}},
\end{equation}%
or%
\begin{equation}
4w\left( 1-z\right) ^{4}-z=0,  \label{BI_algebraic_EOM}
\end{equation}%
where%
\begin{equation}
z=\eta \overline{\eta },~~~~~w=\frac{\phi ^{2}\overline{\phi }^{2}}{\left(
16\Lambda ^{4}\right) ^{2}}.
\end{equation}%
Therefore the normal ordered BI Lagrangian reads%
\begin{eqnarray}
\mathcal{L}_{BI,int}^{NO}\left( \phi ^{2},\overline{\phi }^{2}\right)
&=&8\Lambda ^{4}\frac{\phi ^{2}\overline{\phi }^{2}}{\left( 16\Lambda
^{4}\right) ^{2}}\left( 1+\eta \overline{\eta }\right) \left( 1-\eta 
\overline{\eta }\right) ^{2}  \notag \\
&=&8\Lambda ^{4}w\left( 1+z\right) \left( 1-z\right) ^{2},
\label{BI_normal_ordered_z_w}
\end{eqnarray}%
where $z$ is a solution of (\ref{BI_algebraic_EOM}). This quartic equation
has four solutions, however only one of them is analytic for $w=0$. The
proper solution can be inserted into right hand side of (\ref%
{BI_normal_ordered_z_w}) using the general formula (see also \cite%
{Aschieri:2013aza} where this approach has been used in similar context)%
\begin{equation}
f\left( z_{0}\right) =\frac{1}{2\pi \mathrm{i}}\int_{C}\mathrm{d}z\frac{%
f\left( z\right) }{F\left( z\right) }F^{\prime }\left( z\right)
\label{residue_formula}
\end{equation}%
where $f\left( z\right) $ is an analytic function at $z_{0}$ while $z_{0}$
is a simple zero of $F\left( z\right) $, i.e. 
\begin{equation*}
F\left( z_{0}\right) =0,~~~F^{\prime }\left( z_{0}\right) \neq 0
\end{equation*}%
and there is no other zero of $F\left( z\right) $ inside the closed curve $C$%
. Choosing $f=\mathcal{L}_{BI,int}^{NO}\left( z\right) $ \ and $F\left(
z\right) =z-4w\left( 1-z\right) ^{4}$ we get%
\begin{equation}
\mathcal{L}_{BI,int}^{NO}\left( \phi ^{2},\overline{\phi }^{2}\right)
=8\Lambda ^{4}w\frac{1}{2\pi \mathrm{i}}\oint_{\left\vert z\right\vert
=\varepsilon }\mathrm{d}z\frac{\left( 1+z\right) \left( 1-z\right) ^{2}}{%
z-4w\left( 1-z\right) ^{4}}\left( 1+16w\left( 1-z\right) ^{3}\right) .
\end{equation}%
The contour in the last formula picks up the solution of (\ref%
{BI_algebraic_EOM}) which vanishes for $\phi ,\overline{\phi }\rightarrow 0$
provided $\varepsilon $ is small enough. The integrand can be expanded in
powers of $w$ and we get%
\begin{eqnarray}
\mathcal{L}_{BI,int}^{NO}\left( \phi ^{2},\overline{\phi }^{2}\right)
&=&8\Lambda ^{4}w\frac{1}{2\pi \mathrm{i}}\oint_{\left\vert z\right\vert
=\varepsilon }\frac{\mathrm{d}z}{z}\left( 1+z\right) \left( 1-z\right)
^{2}\left( 1+16w\left( 1-z\right) ^{3}\right)  \notag \\
&&\times \sum_{n=0}^{\infty }4^{n}z^{-n}\left( 1-z\right) ^{4n}w^{n}
\end{eqnarray}%
and after some straightforward algebra%
\begin{eqnarray}
\mathcal{L}_{BI,int}^{NO}\left( \phi ^{2},\overline{\phi }^{2}\right)
&=&8\Lambda ^{4}\sum_{n=0}^{\infty }4^{n}w^{n+1}\frac{1}{2\pi \mathrm{i}}%
\oint_{\left\vert z\right\vert =\varepsilon }\frac{\mathrm{d}z}{z}\left(
1+4z+3z^{2}\right) \\
&&\times \sum_{k=0}^{4n+1}\left( 
\begin{array}{c}
4n+1 \\ 
k%
\end{array}%
\right) \left( -1\right) ^{k}z^{-n+k}
\end{eqnarray}%
Calculating the residue at $z=0$ we get in the end 
\begin{eqnarray*}
&&\frac{1}{2\pi \mathrm{i}}\oint_{\left\vert z\right\vert =\varepsilon }%
\frac{\mathrm{d}z}{z}\left( 1+4z+3z^{2}\right) \sum_{k=0}^{4n+1}\left( 
\begin{array}{c}
4n+1 \\ 
k%
\end{array}%
\right) \left( -1\right) ^{k}z^{-n+k} \\
&=&\left( -1\right) ^{n}\left[ \left( 
\begin{array}{c}
4n+1 \\ 
n%
\end{array}%
\right) -4\left( 
\begin{array}{c}
4n+1 \\ 
n-1%
\end{array}%
\right) +3\left( 
\begin{array}{c}
4n+1 \\ 
n-2%
\end{array}%
\right) \right] \\
&=&\left( -1\right) ^{n}2\frac{\left( 4n+1\right) !}{\left( 3n+2\right)
!\left( n+1\right) !}
\end{eqnarray*}%
and thus%
\begin{equation}
\mathcal{L}_{BI,int}^{NO}\left( \phi ^{2},\overline{\phi }^{2}\right)
=16\Lambda ^{4}w\sum_{n=0}^{\infty }\frac{\left( 4n+1\right) !}{\left(
3n+2\right) !\left( n+1\right) !}\left( -4w\right) ^{n}
\label{L_BI_normal_series}
\end{equation}%
where%
\begin{equation}
w=\frac{\phi ^{2}\overline{\phi }^{2}}{\left( 16\Lambda ^{4}\right) ^{2}}.
\end{equation}%
The power series (\ref{L_BI_normal_series}) can be summed up and $\mathcal{L}%
_{BI,int}^{NO}\left( \phi ^{2},\overline{\phi }^{2}\right) $ is given in a
closed form as%
\begin{equation}
\mathcal{L}_{BI,int}^{NO}\left( \phi ^{2},\overline{\phi }^{2}\right) =-%
\frac{3}{2}\Lambda ^{4}\left\{ _{3}F_{2}\left[ \left( -\frac{1}{2},-\frac{1}{%
4},\frac{1}{4}\right) ,\left( \frac{1}{3},\frac{2}{3}\right) ,-\frac{2^{2}}{%
3^{3}}\frac{\phi ^{2}\overline{\phi }^{2}}{\Lambda ^{8}}\right] -1\right\}.
\label{L_BI_nornal_hypergeometric}
\end{equation}%
Remarkably the same function appears in the expression for the
hypergeometric form of BI Lagrangian found in \cite{Aschieri:2013nda} and 
\cite{Aschieri:2013aza}. Of course this is not an accidental coincidence, as
we have discussed above.

Explicitly we get for the weak field expansion 
\begin{equation}
\mathcal{L}_{BI,int}^{NO}\left( \phi ^{2},\overline{\phi }^{2}\right) =\frac{%
\phi ^{2}\overline{\phi }^{2}}{32\Lambda ^{4}}-\frac{\left[ \phi ^{2}%
\overline{\phi }^{2}\right] ^{2}}{2048\Lambda ^{12}}+3\frac{\left[ \phi ^{2}%
\overline{\phi }^{2}\right] ^{3}}{131072\Lambda ^{20}}+\ldots
\label{L_BI_normal_rdered_explicit}
\end{equation}%
The simple form of (\ref{L_BI_normal_rdered_explicit}) should be compared
with the expansion of the original Lagrangian (\ref{L_BI_original_phi_phibar}%
) 
\begin{eqnarray*}
\mathcal{L}_{BI,int}\left( \phi ^{2},\overline{\phi }^{2}\right) &=&\frac{%
\phi ^{2}\overline{\phi }^{2}}{32\Lambda ^{4}}-\frac{\phi ^{2}\overline{\phi 
}^{2}\left[ \phi ^{2}+\overline{\phi }^{2}\right] }{256\Lambda ^{8}}+\frac{%
\phi ^{2}\overline{\phi }^{2}\left[ \left( \phi ^{2}\right) ^{2}+3\phi ^{2}%
\overline{\phi }^{2}+\left( \overline{\phi }^{2}\right) ^{2}\right] }{%
2048\Lambda ^{12}} \\
&&-\frac{\phi ^{2}\overline{\phi }^{2}\left( \phi ^{2}+\overline{\phi }%
^{2}\right) \left[ \left( \phi ^{2}\right) ^{2}+5\phi ^{2}\overline{\phi }%
^{2}+\left( \overline{\phi }^{2}\right) ^{2}\right] }{16384\Lambda ^{16}} \\
&&+\frac{\phi ^{2}\overline{\phi }^{2}}{131072\Lambda ^{20}}\left[ \left(
\phi ^{2}\right) ^{4}+10\left( \phi ^{2}\right) ^{3}\overline{\phi }%
^{2}\right. \\
&&\left. +20\left( \phi ^{2}\overline{\phi }^{2}\right) ^{2}+10\phi
^{2}\left( \overline{\phi }^{2}\right) ^{3}+\left( \overline{\phi }%
^{2}\right) ^{4}\right] +\ldots
\end{eqnarray*}%
for which the helicity conservation is a result of subtle cancellations of
direct and induced contact terms (i.e. those stemming form gluing together
the original vertices with local parts of the propagators).

\subsection{The simplest helicity conserving theory and the Bossard-Nicolai
model}

Let us now illustrate the inverse problem: suppose that the normal ordered
Lagrangian is known and try to identify the original one. Our example will
be the apparently simplest helicity conserving theory which corresponds to
normal ordered Lagrangian with only one quartic vertex (such a theory was
assumed in \cite{Ivanov:2013jba} as the simplest interaction (SI) model)%
\begin{equation}
\mathcal{L}_{int}^{NO}\left( \phi ^{2},\overline{\phi }^{2}\right) =\frac{%
\lambda }{4}\phi ^{2}\overline{\phi }^{2}.  \label{simplest_NO}
\end{equation}%
In this case we get the formula for the original Lagrangian (\ref%
{L_int_inversion}) in the form%
\begin{eqnarray}
\mathcal{L}_{int}\left( \phi ^{2},\overline{\phi }^{2}\right) &=&\frac{%
\lambda }{4}\varphi ^{2}\overline{\varphi }^{2}+4\varphi ^{2}\left( \frac{%
\lambda }{4}\overline{\varphi }^{2}\right) ^{2}+4\overline{\varphi }%
^{2}\left( \frac{\lambda }{4}\varphi ^{2}\right) ^{2}  \notag \\
&=&\frac{\lambda }{4}\varphi ^{2}\overline{\varphi }^{2}\left[ \left(
1+\lambda \varphi ^{2}\right) \left( 1+\lambda \overline{\varphi }%
^{2}\right) -\lambda ^{2}\varphi ^{2}\overline{\varphi }^{2}\right]
\label{L_BN_phi_phibar}
\end{eqnarray}%
while the algebraic equations determining $\phi ^{2},\overline{\phi }^{2}$
as a functions of $\varphi ^{2}$ and $\overline{\varphi }^{2}$(\ref%
{L_int_inversion_equation}) simplifies to 
\begin{equation}
\varphi ^{2}\left( 1+\lambda \overline{\varphi }^{2}\right) ^{2}-\phi
^{2}=0,~~~~\overline{\varphi }^{2}\left( 1+\lambda \varphi ^{2}\right) ^{2}-%
\overline{\phi }^{2}=0.  \label{L_BN_algebraic}
\end{equation}%
Let us denote for short $z=\varphi ^{2}$ and $\overline{z}=\overline{\varphi 
}^{2}$. The generalization of (\ref{residue_formula}) to the case of two
variables reads in our case%
\begin{equation}
\mathcal{L}_{int}\left( \phi ^{2},\overline{\phi }^{2}\right) =\frac{1}{%
\left( 2\pi \mathrm{i}\right) ^{2}}\oint\limits_{\left\vert z\right\vert
,\left\vert \overline{z}\right\vert =\varepsilon }\frac{\mathrm{d}z\mathrm{d}%
\overline{z}}{h\left( z,\overline{z}\right) \overline{h}\left( z,\overline{z}%
\right) }\det \frac{\partial \left( h,\overline{h}\right) }{\partial \left(
z,\overline{z}\right) }f(z,\overline{z}).
\label{residue_formula_2_variables}
\end{equation}%
where we again choose the double contour in order to pick up the right
solution. In the above formula (see (\ref{L_BN_phi_phibar}))%
\begin{equation}
f(z,\overline{z})=\frac{\lambda }{4}z\overline{z}\left[ \left( 1+\lambda 
\overline{z}\right) \left( 1+\lambda z\right) -\lambda ^{2}z\overline{z}%
\right] ,
\end{equation}%
and (see (\ref{L_BN_algebraic}))%
\begin{equation}
h\left( z,\overline{z}\right) =z\left( 1+\lambda \overline{z}\right)
^{2}-\phi ^{2},~~~~~\overline{h}\left( z,\overline{z}\right) =\overline{z}%
\left( 1+\lambda z\right) ^{2}-\overline{\phi }^{2}.
\end{equation}%
For the Jacobian we have then%
\begin{equation}
\det \frac{\partial \left( h,\overline{h}\right) }{\partial \left( z,%
\overline{z}\right) }=\left( 1+\lambda \overline{z}\right) \left( 1+\lambda
z\right) \left[ \left( 1+\lambda \overline{z}\right) \left( 1+\lambda
z\right) -4\lambda ^{2}z\overline{z}\right] .
\end{equation}%
Expanding $\mathcal{L}_{int}\left( \phi ^{2},\overline{\phi }^{2}\right) $
given by (\ref{residue_formula_2_variables}) in powers of $\phi ^{2}$ and $%
\overline{\phi }^{2}$ we get 
\begin{eqnarray}
\mathcal{L}_{int}\left( \phi ^{2},\overline{\phi }^{2}\right) &=&\frac{1}{%
\left( 2\pi \mathrm{i}\right) ^{2}}\frac{\lambda }{4}\sum_{n,m}\left( \phi
^{2}\right) ^{n}\left( \overline{\phi }^{2}\right) ^{m}  \notag \\
&&\times \oint\limits_{\left\vert z\right\vert ,\left\vert \overline{z}%
\right\vert =\varepsilon }\frac{\mathrm{d}z\mathrm{d}\overline{z}}{%
z^{n}\left( 1+\lambda \overline{z}\right) ^{2n+1}\overline{z}^{m}\left(
1+\lambda z\right) ^{2m+1}}  \notag \\
&&\times \left[ \left( 1+\lambda \overline{z}\right) \left( 1+\lambda
z\right) -4\lambda ^{2}z\overline{z}\right] \left[ \left( 1+\lambda 
\overline{z}\right) \left( 1+\lambda z\right) -\lambda ^{2}z\overline{z}%
\right]  \notag \\
&&
\end{eqnarray}%
and after some algebra the double integral can be rewritten a form of the
linear combination of factorized simple variable integrals%
\begin{eqnarray}
\mathcal{L}_{int}\left( \phi ^{2},\overline{\phi }^{2}\right) &=&\frac{1}{%
\left( 2\pi \mathrm{i}\right) ^{2}}\frac{\lambda }{4}\sum_{n,m}\left( \phi
^{2}\right) ^{n}\left( \overline{\phi }^{2}\right) ^{m}  \notag \\
&&\times \oint\limits_{\left\vert z\right\vert ,\left\vert \overline{z}%
\right\vert =\varepsilon }\mathrm{d}z\mathrm{d}\overline{z}\left[
f_{n,2m-1}\left( z\right) f_{m,2n-1}\left( \overline{z}\right) \right. 
\notag \\
&&\left. -5\lambda ^{2}f_{n-1,2m}\left( z\right) f_{m-1,2n}\left( \overline{z%
}\right) +4\lambda ^{4}f_{n-1,2m}\left( z\right) f_{m-1,2n}\left( \overline{z%
}\right) \right]  \notag \\
&&
\end{eqnarray}%
where%
\begin{equation}
f_{k,l}\left( x\right) =\frac{1}{x^{k}\left( 1+\lambda x\right) ^{l}}.
\end{equation}%
The resulting single variable integrals can be evaluated using residue
theorem%
\begin{eqnarray}
\frac{1}{2\pi \mathrm{i}}\oint\limits_{\left\vert z\right\vert =\varepsilon }%
\mathrm{d}zf_{k,l}\left( z\right) &=&\frac{1}{\left( k-1\right) !}\left(
-\lambda \right) ^{k-1}l(l+1)\ldots \left( l+k-2\right)  \notag \\
&=&\left( -\lambda \right) ^{k-1}\left( 
\begin{array}{c}
l+k-2 \\ 
k-1%
\end{array}%
\right) .
\end{eqnarray}%
As a result%
\begin{equation}
\mathcal{L}_{int}\left( \phi ^{2},\overline{\phi }^{2}\right) =\sum_{n,m\geq
1}c_{nm}\left( \phi ^{2}\right) ^{n}\left( \overline{\phi }^{2}\right) ^{m}
\end{equation}%
where the coefficients are explicitly given as%
\begin{eqnarray}
c_{nm} &=&\left( -1\right) ^{n+m}\frac{\lambda ^{n+m-1}}{4}\left[ \left( 
\begin{array}{c}
n+2m-3 \\ 
n-1%
\end{array}%
\right) \left( 
\begin{array}{c}
m+2n-3 \\ 
m-1%
\end{array}%
\right) \right. \\
&&\left. -5\left( 
\begin{array}{c}
n+2m-3 \\ 
n-2%
\end{array}%
\right) \left( 
\begin{array}{c}
m+2n-3 \\ 
m-2%
\end{array}%
\right) \right.  \notag \\
&&\left. +4\left( 
\begin{array}{c}
n+2m-3 \\ 
n-3%
\end{array}%
\right) \left( 
\begin{array}{c}
m+2n-3 \\ 
m-3%
\end{array}%
\right) \right] .  \notag
\end{eqnarray}%
After a simple rearrangement of the binomial coefficients we get finally the
original Lagrangian corresponding to normal ordered one (\ref{simplest_NO})
in the form%
\begin{equation}
\mathcal{L}_{int}=-\frac{1}{4}\sum_{n,m\geq 1}\frac{(-\lambda )^{n+m-1}}{nm}%
\left( 
\begin{array}{c}
n+2m-2 \\ 
n-1%
\end{array}%
\right) \left( 
\begin{array}{c}
m+2n-2 \\ 
m-1%
\end{array}%
\right) \left( \phi ^{2}\right) ^{n}\left( \overline{\phi }^{2}\right) ^{m}.
\label{bossard_nicolai}
\end{equation}%
For identification of this theory let us express back the variables $\phi
^{2}$ and $\overline{\phi }^{2\text{ }}$in terms of the invariants $\mathcal{%
F}$ and $\mathcal{G}$ (see (\ref{F_G_in_terms_of_phi_phibar}))%
\begin{equation}
\phi ^{2}=4\left( \mathcal{F}-\mathrm{i}\mathcal{G}\right) ,~~~~~\overline{%
\phi }^{2}=4\left( \mathcal{F}+\mathrm{i}\mathcal{G}\right)
\end{equation}%
Fixing now $\lambda =g^{2}/8$ we get from (\ref{bossard_nicolai}) 
\begin{eqnarray}
\mathcal{L}_{int} &=&\frac{1}{2}g^{2}\left( \mathcal{F}^{2}+\mathcal{G}%
^{2}\right) -\frac{1}{2}g^{4}\mathcal{F}\left( \mathcal{F}^{2}+\mathcal{G}%
^{2}\right)  \notag \\
&&+\frac{1}{4}g^{6}\left( \mathcal{F}^{2}+\mathcal{G}^{2}\right) \left( 3%
\mathcal{F}^{2}+\mathcal{G}^{2}\right) -\frac{1}{8}g^{8}\mathcal{F}\left( 
\mathcal{F}^{2}+\mathcal{G}^{2}\right) \left( 11\mathcal{F}^{2}+7\mathcal{G}%
^{2}\right)  \notag \\
&&+\frac{1}{32}g^{10}\left( \mathcal{F}^{2}+\mathcal{G}^{2}\right) \left( 91%
\mathcal{F}^{4}+86\mathcal{F}^{2}\mathcal{G}^{2}+11\mathcal{G}^{4}\right) 
\notag \\
&&-\frac{1}{8}g^{12}\mathcal{F}\left( \mathcal{F}^{2}+\mathcal{G}^{2}\right)
\left( 51\mathcal{F}^{4}+64\mathcal{F}^{2}\mathcal{G}^{2}+17\mathcal{G}%
^{4}\right)  \notag \\
&&+\frac{1}{64}g^{14}\left( \mathcal{F}^{2}+\mathcal{G}^{2}\right) \left( 969%
\mathcal{F}^{6}+1517\mathcal{F}^{4}\mathcal{G}^{2}+623\mathcal{F}^{2}%
\mathcal{G}^{4}+43\mathcal{G}^{6}\right) +\ldots  \notag \\
&&
\end{eqnarray}%
which can be identified with the first seven terms of the expansion of the
interaction Lagrangian of the Bossard-Nicolai model; these terms were
calculated explicitly in \cite{Carrasco:2011jv} using different method. We
can therefore conclude that the Lagrangian (\ref{bossard_nicolai})
corresponds to the BN model.

\subsection{From normal ordering to original Lagrangian - the general case
of self-dual theory}

According to the previous subsections, in the general case, the self-dual
theory is obtained form the manifestly $U\left( 1\right) $ invariant
interaction Lagrangian as%
\begin{equation}
\mathcal{L}_{int}\left( \phi ^{2},\overline{\phi }^{2}\right) =\mathcal{L}%
_{int}^{NO}\left( \varphi ^{2},\overline{\varphi }^{2}\right) +4\varphi
^{2}\left( \frac{\partial \mathcal{L}_{int}^{NO}\left( \varphi ^{2},%
\overline{\varphi }^{2}\right) }{\partial \varphi ^{2}}\right) ^{2}+4%
\overline{\varphi }^{2}\left( \frac{\partial \mathcal{L}_{int}^{NO}\left(
\varphi ^{2},\overline{\varphi }^{2}\right) }{\partial \overline{\varphi }%
^{2}}\right) ^{2},
\end{equation}%
where $\varphi ^{2},\overline{\varphi }^{2}$ are solutions of 
\begin{equation}
\phi ^{2}=\varphi ^{2}\left( 1+4\frac{\partial \mathcal{L}_{int}^{NO}\left(
\varphi ^{2},\overline{\varphi }^{2}\right) }{\partial \varphi ^{2}}\right)
^{2},~~~~\overline{\phi }^{2}=\overline{\varphi }^{2}\left( 1+4\frac{%
\partial \mathcal{L}_{int}^{NO}\left( \varphi ^{2},\overline{\varphi }%
^{2}\right) }{\partial \overline{\varphi }^{2}}\right) ^{2}
\end{equation}%
Note, that duality invariance and Lorentz invariance requires that $\mathcal{%
L}_{int}\left( \varphi ^{2},\overline{\varphi }^{2}\right) $ is function of
the invariant combination $\varphi ^{2}\overline{\varphi }^{2}$ 
\begin{equation}
\mathcal{L}_{int}^{NO}\left( \varphi ^{2},\overline{\varphi }^{2}\right)
\equiv L\left( \varphi ^{2}\overline{\varphi }^{2}\right) 
\end{equation}%
and thus we can write%
\begin{eqnarray}
\mathcal{L}_{int} &=&L+4\varphi ^{2}\left( \overline{\varphi }^{2}L^{\prime
}\right) ^{2}+4\overline{\varphi }^{2}\varphi ^{2}\left( \varphi
^{2}L^{\prime }\right) ^{2} \\
&=&L+4\varphi ^{2}\overline{\varphi }^{2}\left( \overline{\varphi }%
^{2}+\varphi ^{2}\right) L^{\prime 2}  \notag
\end{eqnarray}%
where the prime means a derivative of $L$ with respect to $\varphi ^{2}%
\overline{\varphi }^{2}$. The algebraic equations defining $\varphi ^{2}$
and $\overline{\varphi }^{2}$ in terms of $\phi ^{2}$ and $\overline{\phi }%
^{2}$ are then 
\begin{equation}
\phi ^{2}=\varphi ^{2}\left( 1+4\overline{\varphi }^{2}L^{\prime }\right)
^{2},~~~~\overline{\phi }^{2}=\overline{\varphi }^{2}\left( 1+4\varphi
^{2}L^{\prime }\right) ^{2}.
\end{equation}%
or taking the square root%
\begin{equation}
\sqrt{\phi ^{2}}=\sqrt{\varphi ^{2}}\left( 1+4\overline{\varphi }%
^{2}L^{\prime }\right) ,~~~~\sqrt{\overline{\phi }^{2}}=\sqrt{\overline{%
\varphi }^{2}}\left( 1+4\varphi ^{2}L^{\prime }\right) .
\end{equation}%
Let us introduce new variables%
\begin{equation}
x_{\pm }=\frac{1}{2}\left( \sqrt{\varphi ^{2}}\pm \sqrt{\overline{\varphi }%
^{2}}\right) ,~~~~~~X_{\pm }=\frac{1}{2}\left( \sqrt{\phi ^{2}}\pm \sqrt{%
\overline{\phi }^{2}}\right) ,
\end{equation}%
in terms of which we get%
\begin{eqnarray}
\sqrt{\varphi ^{2}}\sqrt{\overline{\varphi }^{2}}
&=&x_{+}^{2}-x_{-}^{2},~~~~~\varphi ^{2}+\overline{\varphi }^{2}=2\left(
x_{+}^{2}+x_{-}^{2}\right) , \\
X_{\pm } &=&x_{\pm }\left( 1\pm 4\sqrt{\varphi ^{2}}\sqrt{\overline{\varphi }%
^{2}}L^{\prime }\left( \varphi ^{2}\overline{\varphi }^{2}\right) \right) .
\end{eqnarray}%
Let us further abbreviate $z=\sqrt{\varphi ^{2}}\sqrt{\overline{\varphi }^{2}%
}$. The interaction Lagrangian is then expressed in a compact form%
\begin{equation}
\mathcal{L}_{int}=L\left( z^{2}\right) +8z^{2}L^{\prime }\left( z^{2}\right)
^{2}\left( x_{+}^{2}+x_{-}^{2}\right) 
\end{equation}%
where $x_{\pm }$ are solutions of 
\begin{equation}
X_{\pm }=x_{\pm }\left( 1\pm 4zL^{\prime }\left( z^{2}\right) \right) .
\end{equation}%
Note that we do not need to know $x_{\pm }$ individually but only in the
combinations $x_{+}^{2}+x_{-}^{2}$ and $z$. The latter equations imply for
these 
\begin{eqnarray}
x_{+}^{2}+x_{-}^{2} &=&\frac{X_{+}^{2}}{\left( 1+4zL^{\prime }\left(
z^{2}\right) \right) ^{2}}+\frac{X_{-}^{2}}{\left( 1-4zL^{\prime }\left(
z^{2}\right) \right) ^{2}},  \label{xp_xm_equation} \\
z &=&\frac{X_{+}^{2}}{\left( 1+4zL^{\prime }\left( z^{2}\right) \right) ^{2}}%
-\frac{X_{-}^{2}}{\left( 1-4zL^{\prime }\left( z^{2}\right) \right) ^{2}}.
\label{z_equation1}
\end{eqnarray}%
For the interaction Lagrangian we get therefore 
\begin{equation}
\mathcal{L}_{int}=L\left( z^{2}\right) +8z^{2}L^{\prime }\left( z^{2}\right)
^{2}\left[ \frac{X_{+}^{2}}{\left( 1+4zL^{\prime }\left( z^{2}\right)
\right) ^{2}}+\frac{X_{-}^{2}}{\left( 1-4zL^{\prime }\left( z^{2}\right)
\right) ^{2}}\right] ,
\end{equation}%
where $z$ is solution of single equation (\ref{z_equation1}). Using this
equation we can finally simplify $\mathcal{L}_{int}$ to the form%
\begin{equation}
\mathcal{L}_{int}=L\left( z^{2}\right) +2zL^{\prime }\left( z^{2}\right)
\left( \frac{X_{+}^{2}}{1+4zL^{\prime }\left( z^{2}\right) }-\frac{X_{-}^{2}%
}{1-4zL^{\prime }\left( z^{2}\right) }-z\right) .
\label{L_int_general_simplified}
\end{equation}%
Remarkably, we can immediately make sure, that the complete Lagrangian $%
\mathcal{L}=-\frac{1}{4}X_{+}^{2}-\frac{1}{4}X_{-}^{2}+\mathcal{L}_{int}$
represents the solution of the NGZ condition. Indeed, it is an easy exercise
to show that this Lagrangian can be reconstructed according to the general
representation (\ref{p(u)_representation}), (\ref{p(u)_euqation}) with the
identification%
\begin{equation}
p\left( u\right) =\frac{1+4uL^{\prime }\left( u^{2}\right) }{1-4uL^{\prime
}\left( u^{2}\right) },~~~F\left( u\right) =u\left[ 1-16u^{2}L^{\prime
}\left( u^{2}\right) ^{2}\right] ,
\end{equation}%
from which all the other representations discussed in section 3 can be in
principle derived. For instance, the BN model, for which we have $L\left(
z\right) =\frac{\lambda }{4}z$, can be constructed according to (\ref%
{z_equation}) and (\ref{z_Lagrangian}) using the function\footnote{%
See section 3 for passing form $p\left( u\right) $ and $F\left( u\right) $
to $f\left( z\right) $.}%
\begin{equation}
f^{BN}\left( z\right) =4\frac{z}{\lambda }\frac{z-1}{\left( z+1\right) ^{3}}.
\end{equation}

Let us now return to the general case. In order to insert the right solution
of (\ref{z_equation1}) into (\ref{L_int_general_simplified}) we use the same
trick as in the previous subsections and write%
\begin{equation}
\mathcal{L}_{int}=\frac{1}{2\pi \mathrm{i}}\oint\limits_{\left\vert
z\right\vert =\varepsilon }\frac{\mathrm{d}z}{h\left( z\right) }h^{\prime
}\left( z\right) L_{int}\left( z\right) ,
\end{equation}%
where now%
\begin{equation}
h\left( z\right) =z-\left[ \frac{X_{+}^{2}}{\left( 1+4zL^{\prime }\left(
z^{2}\right) \right) ^{2}}-\frac{X_{-}^{2}}{\left( 1-4zL^{\prime }\left(
z^{2}\right) \right) ^{2}}\right] 
\end{equation}%
and $L_{int}\left( z\right) $ is the right hand side of (\ref%
{L_int_general_simplified}). Expanding the integrand in powers of $X_{\pm }$
we get%
\begin{eqnarray}
\mathcal{L}_{int} &=&\frac{1}{2\pi \mathrm{i}}\oint\limits_{\left\vert
z\right\vert =\varepsilon }\frac{\mathrm{d}z}{z}\sum_{n=0}^{\infty }\frac{1}{%
z^{n}}\left[ \frac{X_{+}^{2}}{\left( 1+4zL^{\prime }\left( z^{2}\right)
\right) ^{2}}-\frac{X_{-}^{2}}{\left( 1-4zL^{\prime }\left( z^{2}\right)
\right) ^{2}}\right] ^{n}h^{\prime }\left( z\right) L_{int}\left( z\right)  
\notag \\
&&
\end{eqnarray}%
and finally using the residue theorem%
\begin{eqnarray}
\mathcal{L}_{int} &=&\sum_{n=0}^{\infty }\lim_{z\rightarrow 0}\frac{1}{n!}%
\frac{\mathrm{d}^{n}}{\mathrm{d}z^{n}}\left[ \frac{X_{+}^{2}}{\left(
1+4zL^{\prime }\left( z^{2}\right) \right) ^{2}}-\frac{X_{-}^{2}}{\left(
1-4zL^{\prime }\left( z^{2}\right) \right) ^{2}}\right] ^{n}h^{\prime
}\left( z\right) L_{int}\left( z\right) .  \notag \\
&&
\end{eqnarray}%
The latter formula allows to calculate $\mathcal{L}_{int}$ to any desired
order in $X_{\pm }^{2}=2\left( \mathcal{F}\pm \sqrt{\mathcal{F}^{2}+\mathcal{%
G}^{2}}\right) $ or $\phi ^{2}$ and $\overline{\phi }^{2}$. Writing%
\begin{equation}
L\left( z\right) =\Lambda ^{4}\sum_{n=1}^{\infty }\frac{1}{n!}\frac{c_{4n}}{%
\Lambda ^{8n}}z^{n},
\end{equation}%
where $\Lambda $ is dimensionful scale, we get explicitly%
\begin{eqnarray}
\mathcal{L}_{int} &=&\frac{c_{4}}{\Lambda ^{4}}\phi ^{2}\overline{\phi }%
^{2}-4\frac{c_{4}^{2}}{\Lambda ^{8}}\phi ^{2}\overline{\phi }^{2}\left( \phi
^{2}+\overline{\phi }^{2}\right)   \notag \\
&&+\frac{1}{2\Lambda ^{12}}\phi ^{2}\overline{\phi }^{2}\left\{ 32c_{4}^{3}%
\left[ \left( \phi ^{2}\right) ^{2}+4\phi ^{2}\overline{\phi }^{2}+\left( 
\overline{\phi }^{2}\right) ^{2}\right] +c_{8}\phi ^{2}\overline{\phi }%
^{2}\right\}   \notag \\
&&-8\frac{c_{4}}{\Lambda ^{16}}\phi ^{2}\overline{\phi }^{2}\left( \phi ^{2}+%
\overline{\phi }^{2}\right) \left\{ 8c_{4}^{3}\left[ \left( \phi ^{2}\right)
^{2}+9\phi ^{2}\overline{\phi }^{2}+\left( \overline{\phi }^{2}\right) ^{2}%
\right] +c_{8}\phi ^{2}\overline{\phi }^{2}\right\}   \notag \\
&&+\ldots 
\end{eqnarray}%
As expected, the couplings at individual terms are related. Notice e.g. the
relation between the four-point and six-point interaction. This relation
implies, that any two (analytic) self-dual theories which have the same
four-point interaction (once the coupling constants of the four-point terms
are adjusted appropriately) have also the same six-point vertex. This
explains e.g. the equivalence of the BI and BN models up to $O(F_{\mu \nu
}^{8})$ , which might seem to be an accidental coincidence. It also prevents
any one loop effective Euler-Heisenberg Lagrangian to be self-dual beyond
the four-point interaction term due to the mismatch of the powers of the
fine structure constants at the four-point and six-point vertex\footnote{%
Similar observation was made already in \cite{Hagiwara:1980jg}, where the
matching of the BI and Euler Heisenberg Lagrangian was discussed.}.

\subsection{Normal ordered form of implicitly defined self-dual Lagrangians}

In the previous subsection we have mentioned that, once the normal ordered
interaction Lagrangian $\mathcal{L}_{int}^{NO}\left( \phi ^{2},\overline{%
\phi }^{2}\right) =L\left( \phi ^{2}\overline{\phi }^{2}\right) $ for the
self-dual theory is known, we can at least in principle construct the
original Lagrangian using the formula for the general solution of the NGZ
condition (\ref{p(u)_representation}), (\ref{p(u)_euqation}) with the
identification%
\begin{equation}
p\left( u\right) =\frac{1+4uL^{\prime }\left( u^{2}\right) }{1-4uL^{\prime
}\left( u^{2}\right) },~~~F\left( u\right) =u\left[ 1-16u^{2}L^{\prime
}\left( u^{2}\right) ^{2}\right] .  \label{p(u)F(u)}
\end{equation}%
Quite remarkably, this relation works also in the reversed direction.
Suppose e.g. that the solution of the NGZ identity is given by equations (%
\ref{z_equation}) and (\ref{z_Lagrangian}) with the known function $f\left(
z\right) $ and let us derive the normal ordered Lagrangian directly from
this function. Let us rewrite the first equation of (\ref{p(u)F(u)}) as%
\begin{equation}
4uL^{\prime }\left( u^{2}\right) =\frac{p-1}{p+1},  \label{Lprime_p_relarion}
\end{equation}%
and suppose it can be solved in order to express $u$ as a function of $p$.
Using now the identification $F\left( u\right) =f\left( p\left( u\right)
\right) $ (see (\ref{p(u)_euqation}) and (\ref{F(u)_definition})), the
second relation of (\ref{p(u)F(u)}) can be rewritten in terms of the
variable $p$ 
\begin{equation}
f\left( p\right) =u\left( p\right) \left[ 1-\left( \frac{p-1}{p+1}\right)
^{2}\right] .
\end{equation}%
The above solution $u\left( p\right) $ has to be therefore given by%
\begin{equation}
u\left( p\right) =f\left( p\right) \frac{\left( p+1\right) ^{2}}{4p}.
\label{u(p)}
\end{equation}%
Inserting this back into (\ref{Lprime_p_relarion}) and multiplying by $%
u^{\prime }\left( p\right) $ given explicitly by (\ref{u(p)}) we get%
\begin{equation}
2u^{\prime }\left( p\right) u\left( p\right) L^{\prime }\left( u\left(
p\right) ^{2}\right) =\frac{1}{2}u^{\prime }\left( p\right) \frac{p-1}{p+1}
\end{equation}%
where the right hand side is now known. Finally, up to an inessential
constant\ 
\begin{equation}
L\left( \phi ^{2}\overline{\phi }^{2}\right) =\frac{1}{2}\int \mathrm{d}p%
\frac{p-1}{p+1}\frac{\mathrm{d}}{\mathrm{d}p}\left[ f\left( p\right) \frac{%
\left( p+1\right) ^{2}}{4p}\right] |_{p=p\left( \phi ^{2}\overline{\phi }%
^{2}\right) }  \label{L_integral_over_p}
\end{equation}%
where $p\left( \phi ^{2}\overline{\phi }^{2}\right) $ is a solution of
(\ref{u(p)}) written in the form 
\begin{equation}
\sqrt{\phi ^{2}\overline{\phi }^{2}}=f\left( p\right) \frac{\left(
p+1\right) ^{2}}{4p},
\end{equation}%
with respect to $p$. \ Of course, to get the normal ordered interaction
Lagrangian in a closed form we have to be able to solve the latter equation
explicitly.

Let us give a simple example of the application of this general
prescription. Take a solution of the NGZ condition in the form (\ref%
{z_equation}) and (\ref{z_Lagrangian}) with%
\begin{equation}
f\left( z\right) =4\Lambda ^{4}\sqrt{z}\frac{1-z}{\left( 1+z\right) ^{2}},
\end{equation}%
which satisfies the analyticity condition (\ref{f_symmetry}). Note however,
that the closed form of this solution is not accessible since the equation (%
\ref{z_equation}) is the eight order polynomial equation for $z$. Inserting
this function into (\ref{L_integral_over_p}) we get%
\begin{equation}
L\left( \phi ^{2}\overline{\phi }^{2}\right) =-\frac{\Lambda ^{4}}{2}\left( 
\sqrt{p}+\frac{1}{\sqrt{p}}\right) +C,
\end{equation}%
where $C$ is an integration constant and $p$ is a solution of 
\begin{equation}
\sqrt{\phi ^{2}\overline{\phi }^{2}}=\Lambda ^{4}\left( \frac{1}{\sqrt{p}}-%
\sqrt{p}\right)
\end{equation}%
or explicitly%
\begin{eqnarray}
p\left( \phi ^{2}\overline{\phi }^{2}\right) &=&1+\frac{\phi ^{2}\overline{%
\phi }^{2}}{2\Lambda ^{8}}\pm \sqrt{\frac{\phi ^{2}\overline{\phi }^{2}}{%
\Lambda ^{8}}+\left( \frac{\phi ^{2}\overline{\phi }^{2}}{2\Lambda ^{8}}%
\right) ^{2}}  \notag \\
&=&\left[ 1+\frac{\phi ^{2}\overline{\phi }^{2}}{2\Lambda ^{8}}\mp \sqrt{%
\frac{\phi ^{2}\overline{\phi }^{2}}{\Lambda ^{8}}+\left( \frac{\phi ^{2}%
\overline{\phi }^{2}}{2\Lambda ^{8}}\right) ^{2}}\right] ^{-1}
\end{eqnarray}%
Finally we get for the normal ordered interaction Lagrangian $\mathcal{L}%
_{int}^{NO}\left( \phi ^{2},\overline{\phi }^{2}\right) =L\left( \phi ^{2}%
\overline{\phi }^{2}\right) $%
\begin{eqnarray}
\mathcal{L}_{int}^{NO}\left( \phi ^{2},\overline{\phi }^{2}\right)
&=&\Lambda ^{4}-\frac{\Lambda ^{4}}{2}\sqrt{1+\frac{\phi ^{2}\overline{\phi }%
^{2}}{2\Lambda ^{8}}+\sqrt{\frac{\phi ^{2}\overline{\phi }^{2}}{\Lambda ^{8}}%
+\left( \frac{\phi ^{2}\overline{\phi }^{2}}{2\Lambda ^{8}}\right) ^{2}}} 
\notag \\
&&-\frac{\Lambda ^{4}}{2}\sqrt{1+\frac{\phi ^{2}\overline{\phi }^{2}}{%
2\Lambda ^{8}}-\sqrt{\frac{\phi ^{2}\overline{\phi }^{2}}{\Lambda ^{8}}%
+\left( \frac{\phi ^{2}\overline{\phi }^{2}}{2\Lambda ^{8}}\right) ^{2}}}
\end{eqnarray}%
where we adjusted the integration constant to get $L\left( 0\right) =0$.
Therefore, although the original Lagrangian is not known in a closed form,
we have enough information on the model e.g. for calculation of the
scattering amplitudes using the known normal ordered Lagrangian and the
modified Feynman rules.

\section{Summary and conclusion}

In this paper, we presented a general proof of the equivalence of two
apparently disconnected aspect of the models of nonlinear quantum
electrodynamics, namely the classical duality invariance of the field
equations, which is expressed on the Lagrangian level by the NGZ condition,
and the helicity conservation of the tree level amplitudes. We have shown,
that the tree level S-matrix is invariant with respect to the $U(1)$
rotational symmetry, which expresses the helicity conservation, if and only
if the Lagrangian of the theory satisfies the NGZ conditions. On the level
of the traditional Feynman rules, the helicity conservation is a result of
subtle cancellations between contributions of different Feynman graphs and
as such is far from being manifest. Using a reorganization of the
perturbative calculation by means of generalized normal ordering of the
Lagrangian and introducing corresponding modification of the Feynman rules
we have shown that for the self-dual models the helicity conservation can be
made manifest on the level of individual Feynman graphs. The general
arguments follow two steps: first we have proved that the normal ordered
Lagrangian is invariant with respect to the $U(1)$ rotational symmetry if
and only if the NGZ identity for the original Lagrangian is satisfied and
then we have shown that the modified Feynman rules manifestly respect this
symmetry. This allows us to enlarge the above statement on helicity
conservation also to higher loops.

The transformation leading from the original Lagrangian to the normal
ordered one and vice versa can be reformulated as a calculation of the tree
level functional integral over auxiliary fields, i.e. as a substitution of
solutions of the classical equation of motions, which become algebraic
(generally transcendental), into a generating Lagrangian. This enables us to
identify the normal ordered Lagrangian with the off-shell $U(1)$ invariant
interaction part of the auxiliary field Lagrangian developed by Ivanov and
Zupnik \cite{Ivanov:2003uj}, \cite{Ivanov:2012bq} (and with its equivalent
within the approach of Carrasco, Kallosh and Roiban in \cite{Carrasco:2011jv}%
). This gives the latter constructions of the self-dual Lagrangians a clear
physical interpretation.

We have also discussed several aspects of the generalized normal ordering.
Namely we gave a general formula for the coefficients of the weak field
expansion of the original Lagrangian of the self-dual theory provided the
normal ordered Lagrangian is known and we also find the general prescription
for the normal ordered Lagrangian derived form the implicit representation
of the general solution of the NGZ condition.

As an illustration of the above concepts we have calculated two explicit
examples. Namely, as the first one we have found the normal ordered form of
the BI Lagrangian and recovered in this way the hypergeometric form of this
theory presented in \cite{Aschieri:2013nda} and \cite{Aschieri:2013aza}. As
the second example we gave two new representations of the BN model. The
first one corresponds to the implicit construction of the general solution
of the GNZ condition for which we found the generating function $%
f^{BN}\left( z\right) $. As the second one we calculated explicitly the
Lagrangian of the BN model in a form of weak field expansion with explicitly
known coefficients.

\bigskip \textbf{Acknowledgment} The author would like to thank Karol Kampf and Francesco Riva for discussions.  This work is supported in part by
Czech Government projects GA\v{C}R 18-17224S and LTAUSA17069. \bigskip

\bibliographystyle{utphys}
\bibliography{duality}
{}

\end{document}